\documentclass[useAMS,usenatbib,usedcolumn]{mn2e}
\usepackage{epsfig}
\usepackage{amsmath}
\usepackage{amssymb}
\usepackage{mathrsfs}

\title[Magnification bias in the BoRG survey]{A spectroscopically confirmed $z=1.327$ galaxy-scale deflector magnifying a $z\sim8$ Lyman-Break galaxy in the Brightest of Reionizing Galaxies survey}
    
\author[Barone-Nugent et al.]{R. L. Barone-Nugent,$^{1,\star}$ A. Sonnenfeld,$^{2}$ J. S. B. Wyithe,$^{1}$ M. Trenti,$^{1}$ T. Treu,$^{2}$ \newauthor K. B. Schmidt,$^{3}$ P. A. Oesch,$^{4}$ L. Bradley,$^{5}$ and T. Puzia$^{6}$\\
$^{1}$ School of Physics, University of Melbourne, Parkville 3010, VIC, Australia\\
$^{2}$ Department of Physics and Astronomy, University of California, Los Angeles, CA 90095-1547, USA\\
$^{3}$ Department of Physics, University of California, Santa Barbara, CA 93106-9530, USA \\
$^{4}$ Yale Center for Astronomy and Astrophysics, Yale University, New Haven, CT 06520, USA\\
$^{5}$ Space Telescope Science Institute, 3700 San Martin Drive, Baltimore, MD 21218 USA\\
$^{6}$ Institute of Astrophysics, Pontificia Universidad Catolica de Chile, Avenida Vicuna Mackenna 4860, Macul, 7820436, Santiago, Chile\\
$\star$Email: robertbn@student.unimelb.edu.au}

\begin{document}
\date{\today}

\maketitle

\label{firstpage}

\begin{abstract}
We present a detailed analysis of an individual case of gravitational lensing of a $z\sim8$ Lyman-Break galaxy (LBG) in a blank field, identified in Hubble Space Telescope imaging obtained as part of the Brightest of Reionizing Galaxies survey. To investigate the close proximity of the bright ($m_{AB}=25.8$) $Y_{098}$-dropout to a small group of foreground galaxies, we obtained deep spectroscopy of the dropout and two foreground galaxies using VLT/X-Shooter. We detect H-$\alpha$, H-$\beta$, [OIII] and [OII] emission in the brightest two foreground galaxies (unresolved at the natural seeing of $0.8$ arcsec), placing the pair at $z=1.327$. We can rule out emission lines contributing all of the observed broadband flux in $H_{160}$ band at $70\sigma$, allowing us to exclude the $z\sim8$ candidate as a low redshift interloper with broadband photometry dominated by strong emission lines.
The foreground galaxy pair lies at the peak of the luminosity, redshift and separation distributions for deflectors of strongly lensed $z\sim8$ objects, and we make a marginal detection of a demagnified secondary image in the deepest ($J_{125}$) filter.
We show that the configuration can be accurately modelled by a singular isothermal ellipsoidal deflector and a S\'{e}rsic source magnified by a factor of $\mu=4.3\pm0.2$. The reconstructed source in the best-fitting model is consistent with luminosities and morphologies of $z\sim8$ LBGs in the literature.
The lens model yields a group mass of $9.62\pm0.31\times10^{11} M_{\odot}$ and a stellar mass-to-light ratio for the brightest deflector galaxy of $M_{\star}/L_{B}=2.3^{+0.8}_{-0.6} M_{\odot}/L_{\odot}$ within its effective radius. The foreground galaxies' redshifts would make this one of the few strong lensing deflectors discovered at $z>1$.
Deeper imaging would allow for confirmation of the existence of the secondary image and elongation in the primary image, verifying multiple imaging and producing more robust estimations of the image magnifications. 
\end{abstract}
\begin{keywords}
galaxies: high-redshift --- cosmology: observations --- gravitational lensing: strong --- galaxies: luminosity function
\end{keywords}

\section{Introduction}
\indent The epoch of reionisation at redshift $z\sim7-12$ \citep{komatsu2010seven,shull2012critical, ade2013planck} is a key period in the evolution of the Universe when the first galaxies formed and reionised the neutral Hydrogen in the intergalactic medium (IGM). Due to the absorption of emitted photons with energies high enough to ionise neutral Hydrogen, the spectra of these early galaxies have highly attenuated regions bluer than the Lyman-alpha break at $1216\textrm{\AA}$, enabling their identification from broad band imaging with the Lyman-break, or `dropout', technique \citep{steidel1996spectroscopy, giavalisco2002lyman}. This technique has been used successfully to build large galaxy samples out to $z\sim8$, with sources selected using observations from both ground and space based observatories \citep[e.g.][]{bouwens2015uv, finkelstein2014evolution, schmidt2014luminosity, bradley2012brightest, mclure2013new}.\\
\indent These new observations have opened the possibility of quantifying and studying the galaxy luminosity function during the epoch of reionisation as well as its evolution with redshift \citep{khochfar2007evolving, bouwens2008z, castellano2010bright, trenti2010well, bouwens2011ultraviolet, finkelstein2012candels, bradley2012brightest, oesch2012bright, schenker2013uv, robertson2013new, mclure2013new, finkelstein2014evolution, schmidt2014luminosity, bouwens2015uv, robertson2015cosmic}. Measurement of the luminosity function is a key requirement for understanding the production of ionising photons at high redshift, which completes hydrogen reionisation. In this context, it is of fundamental importance to discriminate between intrinsic evolution of the galaxy luminosity function and observational selection effects.\\
\indent One possible effect is gravitational lensing magnification bias \citep[e.g.,][]{wallington1993influence}. While the fraction of random lines of sight at $z\gtrsim6$ that are strongly gravitationally lensed by massive foreground galaxies is around $\sim0.5$ per cent \citep{barkana2000high, comerford2002constraining, wyithe2011distortion, barone2015impact, mason2015correcting}, the steep nature of the high redshift luminosity function results in observationally bright Lyman-Break galaxies (LBGs) having an excess likelihood of having been gravitationally lensed \citep{wyithe2011distortion, fialkov2015distortion}. In general, magnification bias depends on the slope of the luminosity function of the source galaxies: steeper luminosity functions yield more lensed sources because the loss in solid angle is compensated by the steep rise in number counts at deeper flux limits. Therefore, this effect becomes especially important for the bright end of the galaxy luminosity function ($L>L_{\star}$), where $L_{\star}$ is the ``knee'' of the luminosity function in the Schechter parameterization, $\phi(L)dL = \phi_{\star}(L/L_{\star})^{\alpha} \exp^{-(L/L_{\star})}d(L/L_{\star})$. In \citet{barone2015impact} we showed that magnification bias exists in LBG surveys at $\gtrsim4$, and becomes a significant factor in surveys of LBGs at $z\gtrsim7$, with $\sim10$ per cent of LBGs brighter than $L_{\star}$ strongly-lensed in fields without massive structures such as the Hubble eXtreme Deep Field \citep[XDF][]{illingworth2013hst}, and Cosmic Assembly Near-infrared Deep Extragalactic Legacy Survey fields \citep[CANDELS][]{grogin2011candels, koekemoer2011candels}.\\
\indent The results in \citet{barone2015impact} confirm the prediction of \citet{wyithe2011distortion} that $\sim10$ per cent of $z\sim8$ LBGs in a survey with a flux limit around $M_{\star}$, such as the Brightest of Reionizing Galaxies (BoRG) WFC3 HST survey \citep{trenti2012overdensities, bradley2012brightest}, could be strongly lensed. Hence the BoRG survey offers the perfect dataset to extend the findings of \citet{barone2015impact}. In fact, an analysis of the BoRG survey by \cite{mason2015correcting} found that $3-15$ per cent of LBGs in the BoRG fields are expected to be strongly lensed, depending on the specific flux limit of each field. They also investigated individual $z\sim8$ candidates in the BoRG survey and found one good candidate for strong gravitational lensing, having an estimated magnification of $\mu=2.02\pm0.52$. They also identified a further three candidates with lower magnifications of $\mu=1.80\pm0.33$, $\mu=1.54\pm0.62$ and $\mu=1.47\pm0.30$.\\
\indent The relatively small number of $z\sim8$ candidates in the BoRG survey do not allow a statistical approach, such as that undertaken by \citet{barone2015impact}, to detect evidence of gravitational lensing. In this paper, we adopt a direct approach to search for evidence of gravitational lensing by looking at individual LBGs. We only consider the highest signal-to-noise (S/N) LBGs in the BoRG sample, which includes 10 candidates detected at  S/N $\geq8\sigma$ and $m\lesssim27$ ($\approx M_{\star}$ at $z\sim8$). The results from \citet{barone2015impact} and \cite{mason2015correcting} imply that $\approx1$ of these LBGs should be strongly gravitationally lensed. We inspected these 10 LBGs searching for potential gravitationally lensed configurations, and find one $z\sim8$ candidate in close projection to a foreground group of galaxies.\\
\indent This paper is structured as follows; We discuss the data, photometry and spectroscopy in Section \ref{section:data}. In Section \ref{section:MCMC} we introduce the Markov-Chain Monte Carlo (MCMC) method that we have applied to investigate the lensing hypothesis, as well as the MCMC results. In Section \ref{section:comparison} we compare the inferred $z\sim8$ source properties to those in the literature. In Section \ref{section:M/L} we investigate the properties of the deflector. We discuss the evidence for lensing in Section \ref{section:discussion}, and we conclude in Section \ref{section:conclusion}. Throughout this paper we assume a $\Lambda$CDM cosmology of $\Omega_{M}=0.27$, $\Omega_{\Lambda}=0.73$ and $H_{0} = 70$kms$^{-1}$Mpc$^{-1}$. We quote magnitudes in the AB system \citep{oke1983secondary}. We refer to the F600LP, F098M, F125W and F160W HST filters as $V_{600}$, $Y_{098}$, $J_{125}$ and $H_{160}$. \\
%

\section{The search for gravitational lensing in the Brightest of Reionizing Galaxies survey}
\label{section:data}
\subsection{The BoRG survey}
\indent The BoRG survey is a Hubble Wide-Field Camera 3 (WFC3) large, multi-year pure-parallel program aimed at identifying galaxy candidates at $z\gtrsim 7.5$ from their colors based on the Lyman-break technique \citep{steidel1996spectroscopy}. Until recently, the main focus of BoRG has been identification of galaxies at $z\sim 8$ as Y-band dropouts, while ongoing observations (cycle 22; \#13767 PI Trenti) are optimized for selecting $z\sim 8-9$ candidates.

In this work, we use data from observations acquired before cycle 22 of the BoRG program. The core of this earlier dataset is based on WFC3 imaging in four filters ($V_{600}$, $Y_{098}$, $J_{125}$, $H_{160}$) obtained as part of HST programs \#11700 and \#12572 (PI Trenti), complemented by archival data with a similar observational design and the same infrared filters, but with F600LP substituting F606W (HST program \# 11702, PI Yan, \citealt{yan2011probing}). A full description of the survey and of its findings is reported by \citet{trenti2011brightest, trenti2012overdensities}, \citet{bradley2012brightest} and \citet{schmidt2014luminosity}. To summarize, the most recent  comprehensive data release of the BoRG survey yielded $n=38$ $z\sim8$ LBG candidates identified as $Y_{098}$-dropouts with $J_{125}$ magnitudes between $25.5$ and $27.6$ mag \citep{schmidt2014luminosity}. These candidates were obtained by analyzing data over a total area of $350$ arcmin$^{2}$ divided in $71$ independent lines of sight.

\subsection{Search for a candidate strongly lensed $z\sim8$ LBG}
\begin{figure*}
\includegraphics[trim=0 30 50 20, clip, scale=.175]{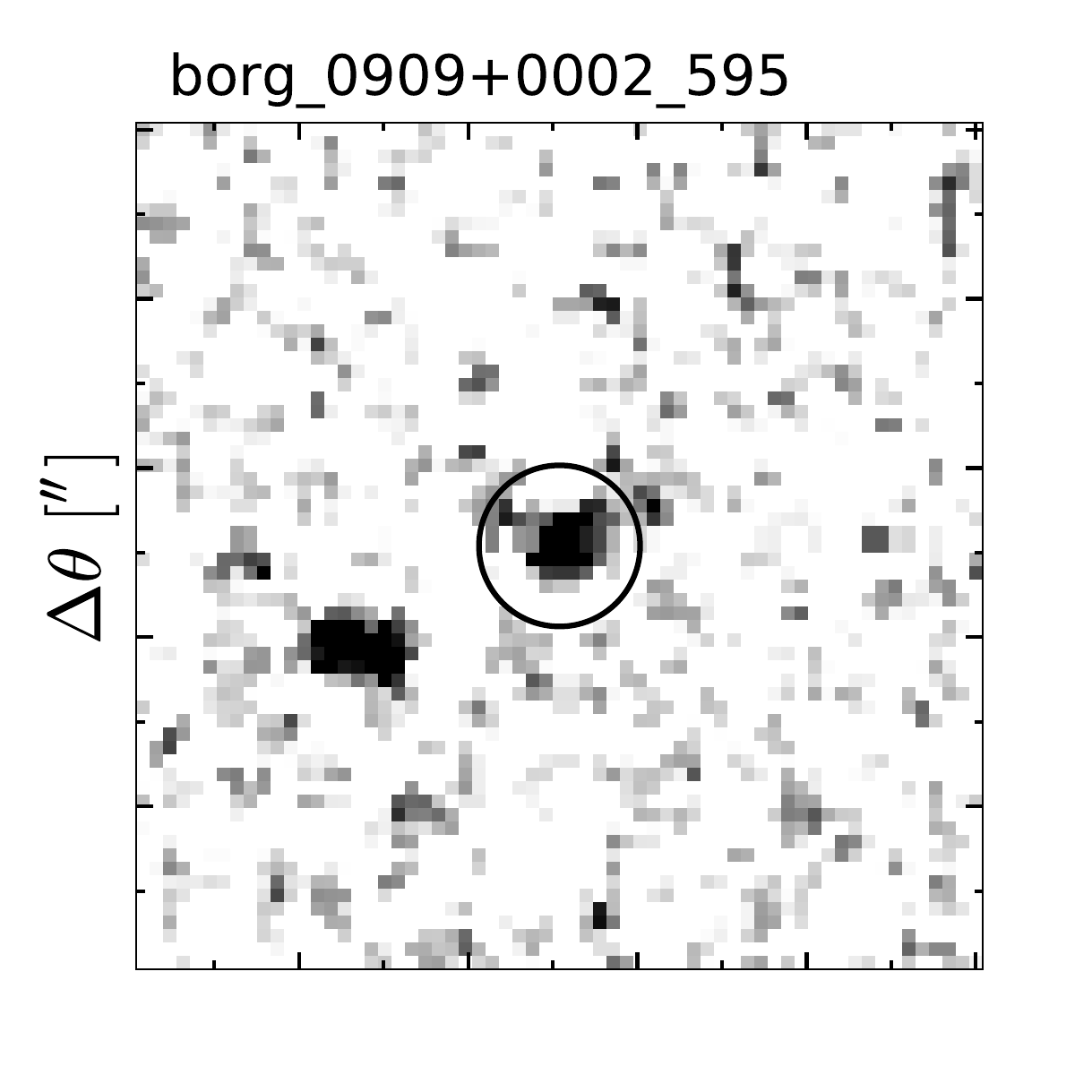}
\includegraphics[trim=0 30 50 20, clip, scale=.175]{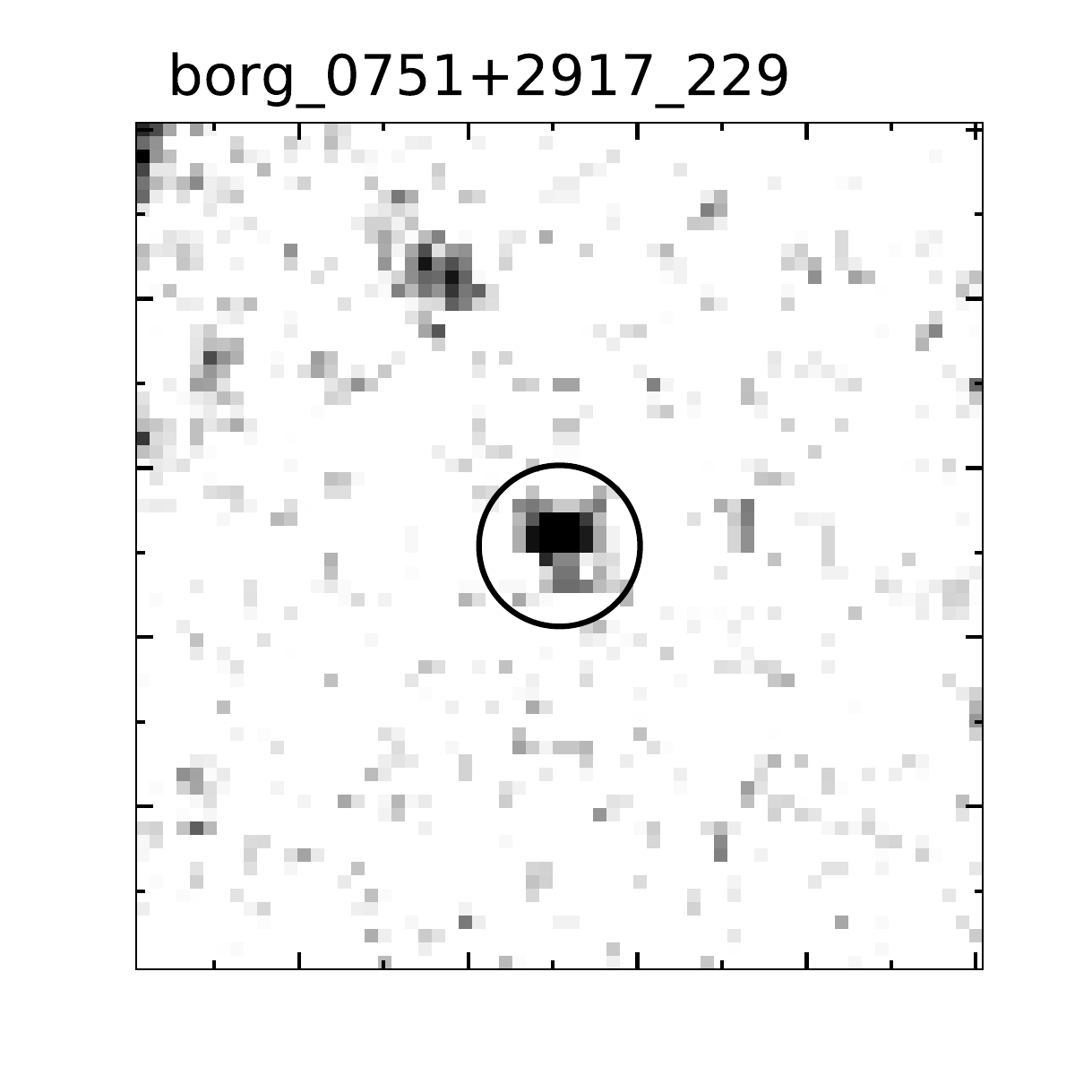}
\includegraphics[trim=0 30 50 20, clip, scale=.175]{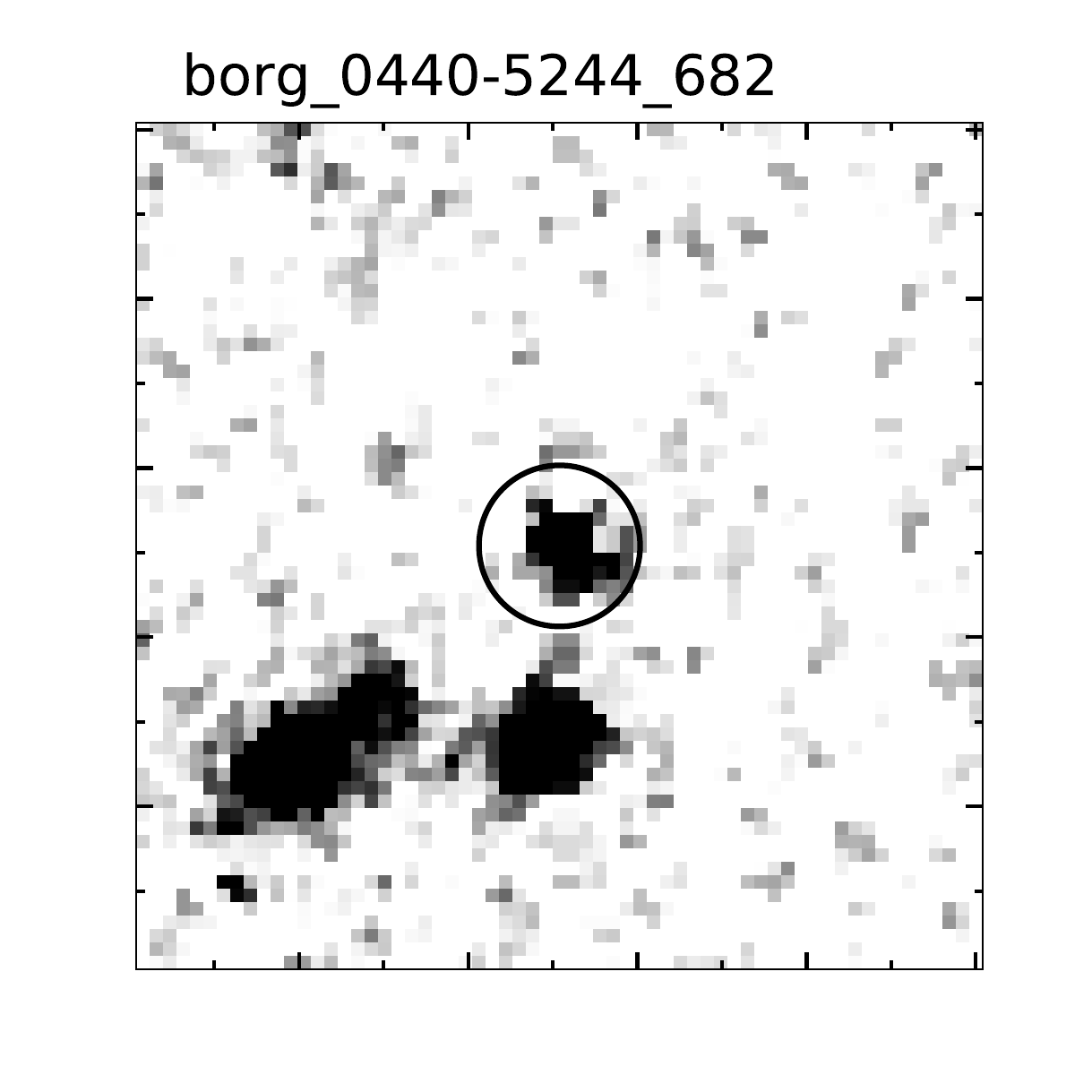}
\includegraphics[trim=0 30 50 20, clip, scale=.175]{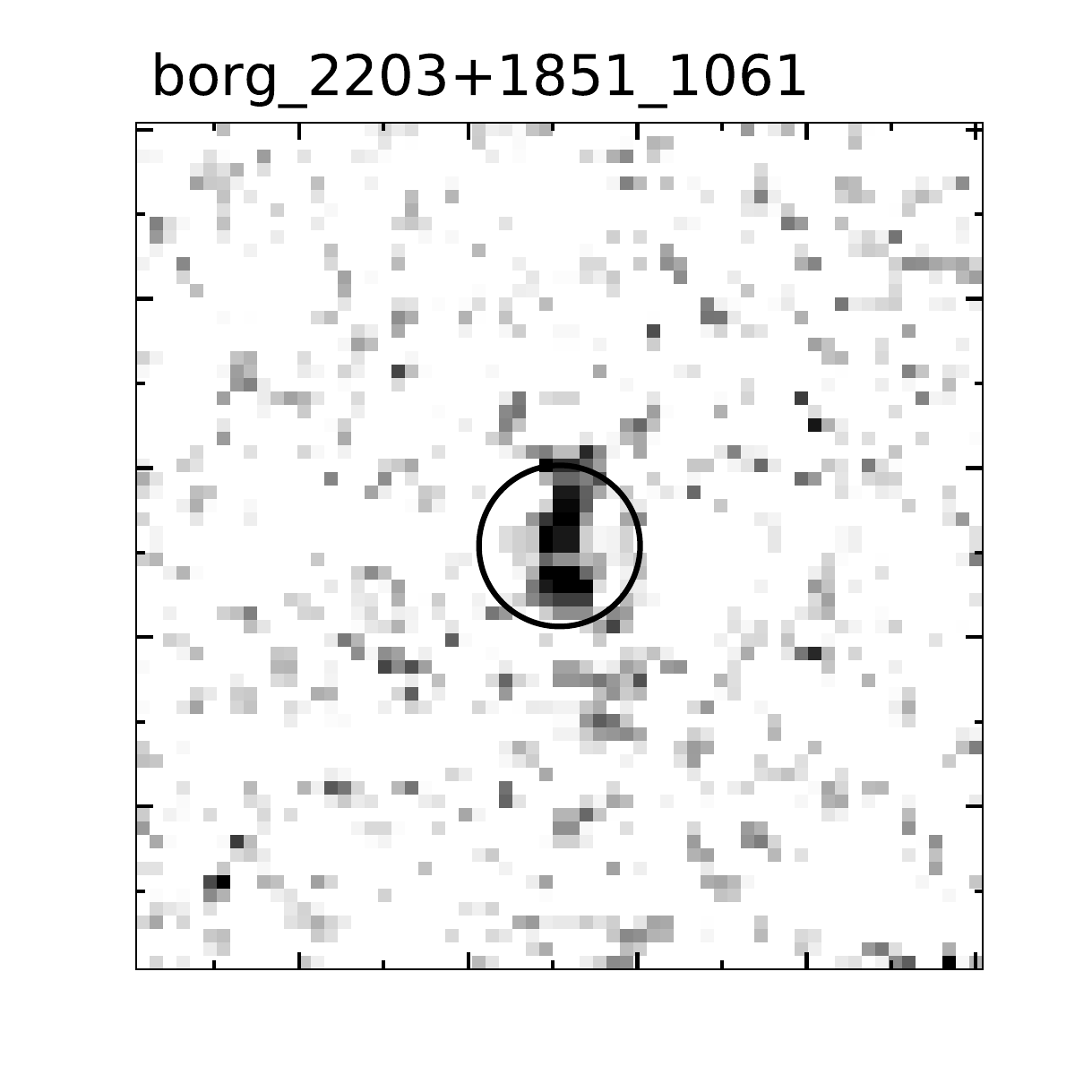}
\includegraphics[trim=0 30 50 20, clip, scale=.175]{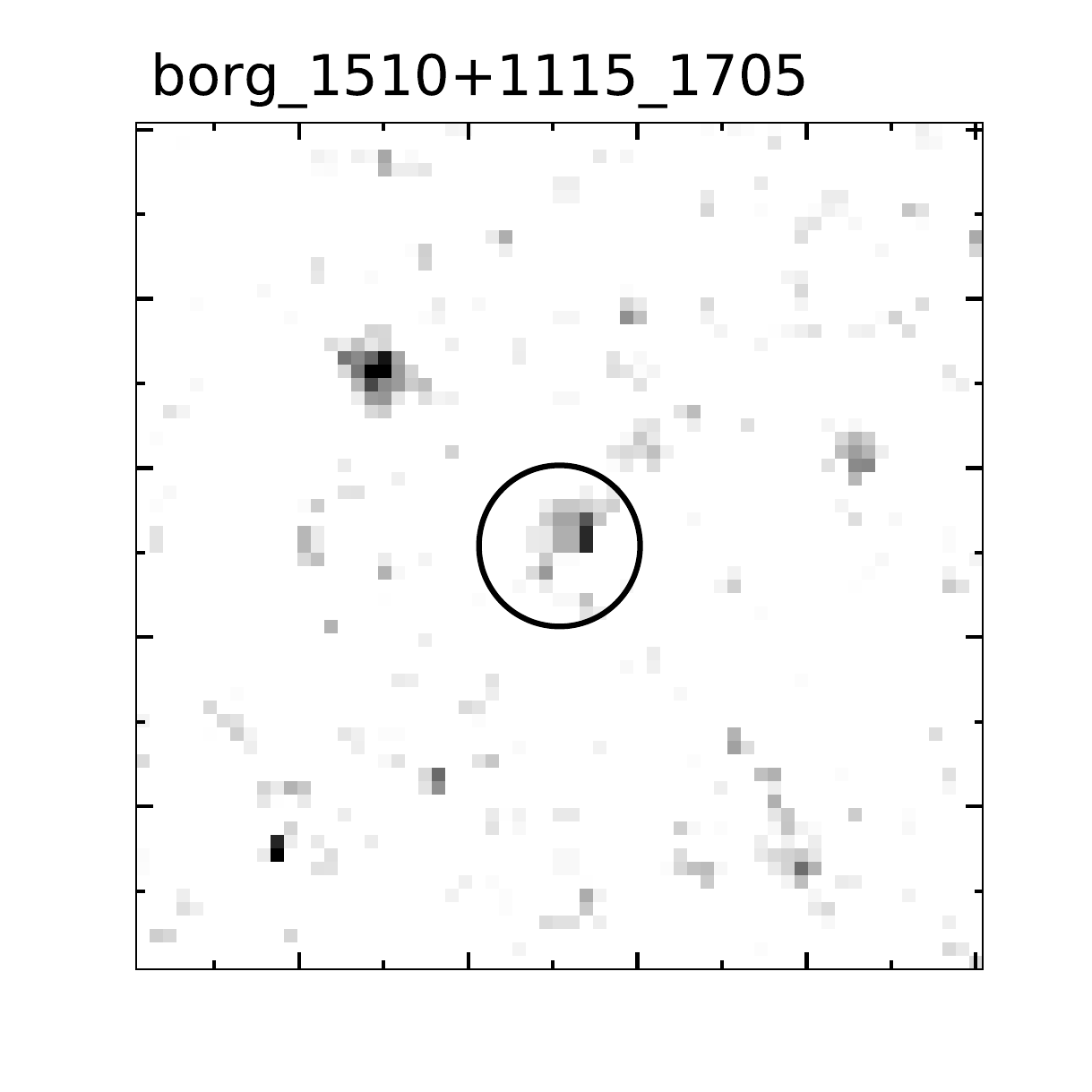}\\
\includegraphics[trim=0 15 50 0, clip, scale=.175]{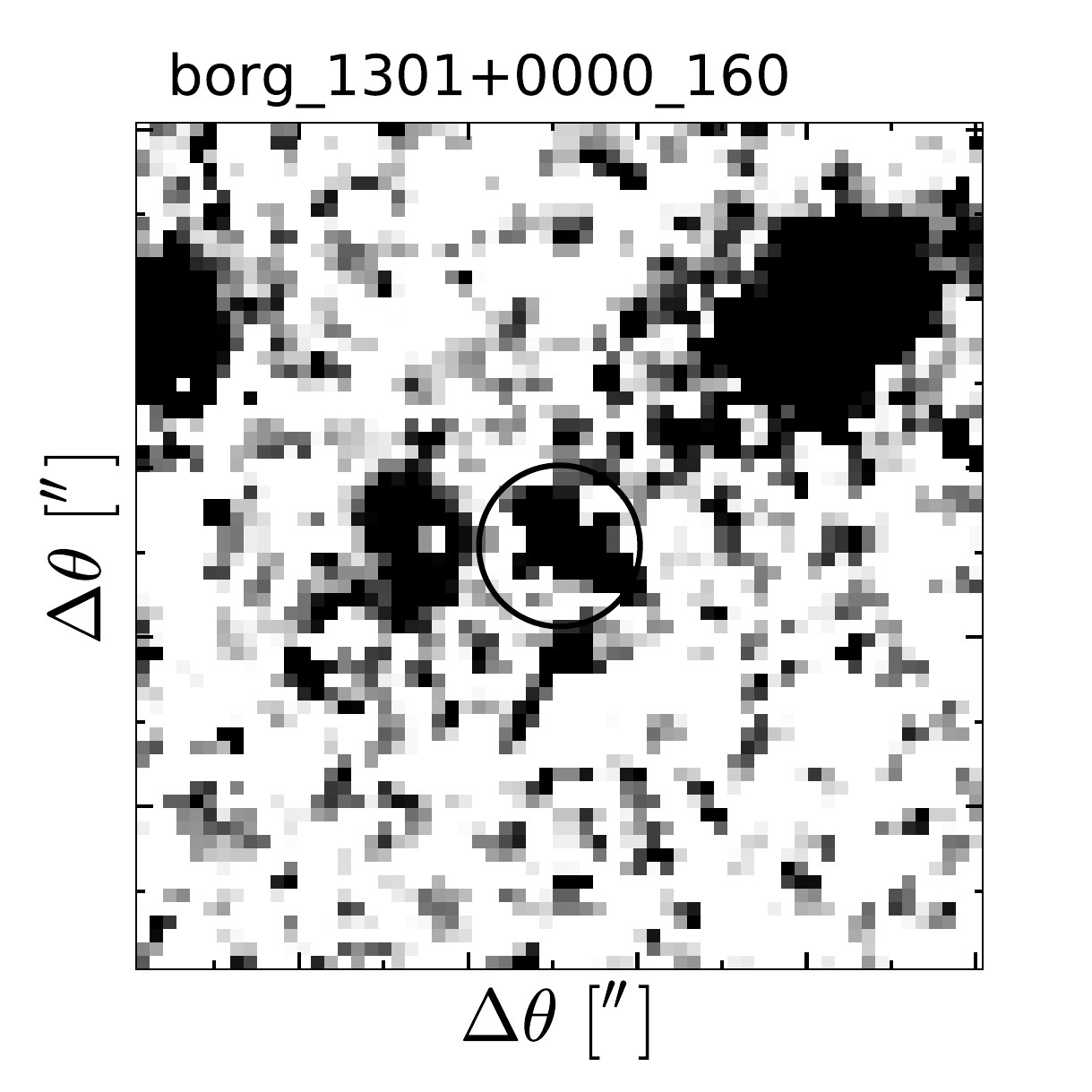}
\includegraphics[trim=0 15 50 0, clip, scale=.175]{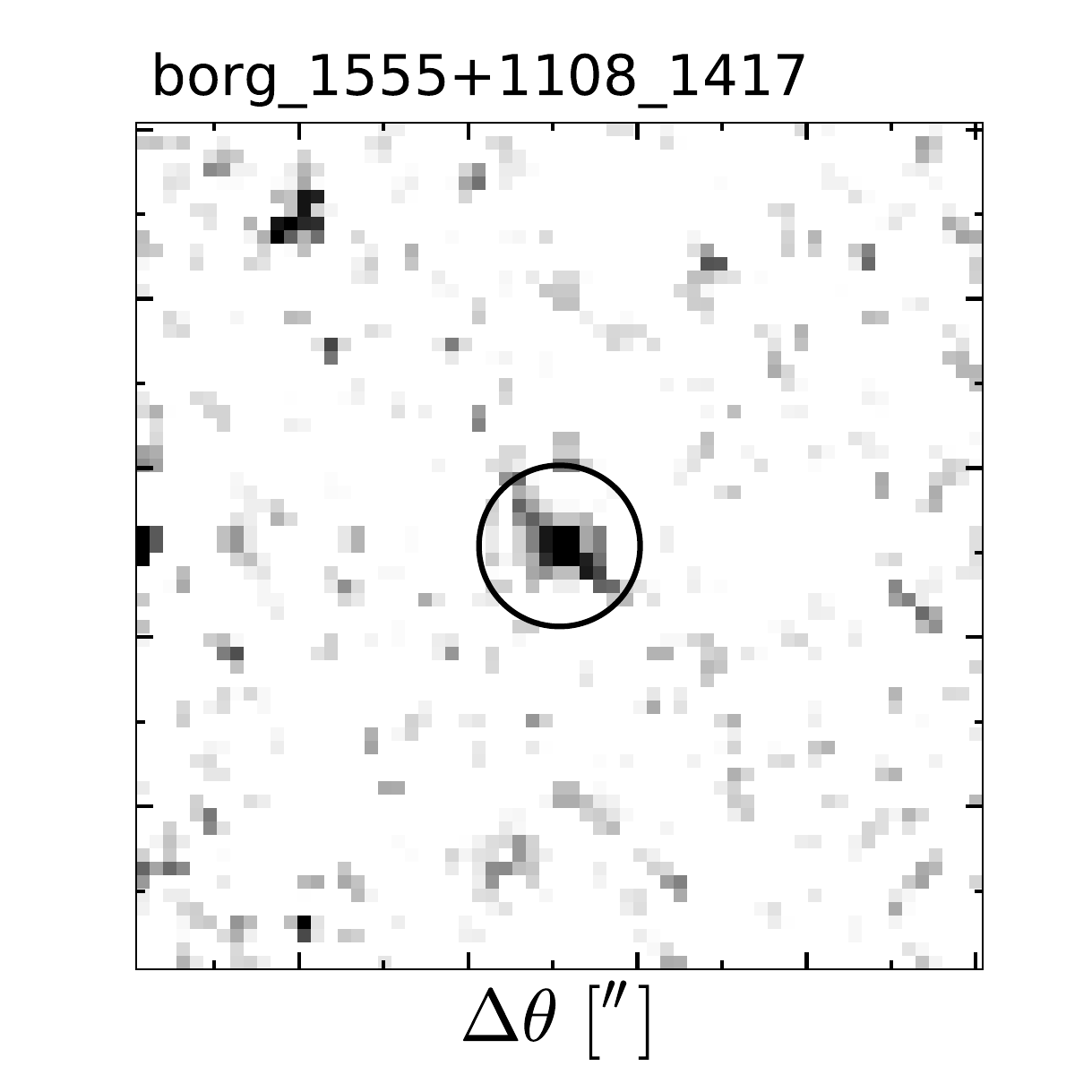}
\includegraphics[trim=0 15 50 0, clip, scale=.175]{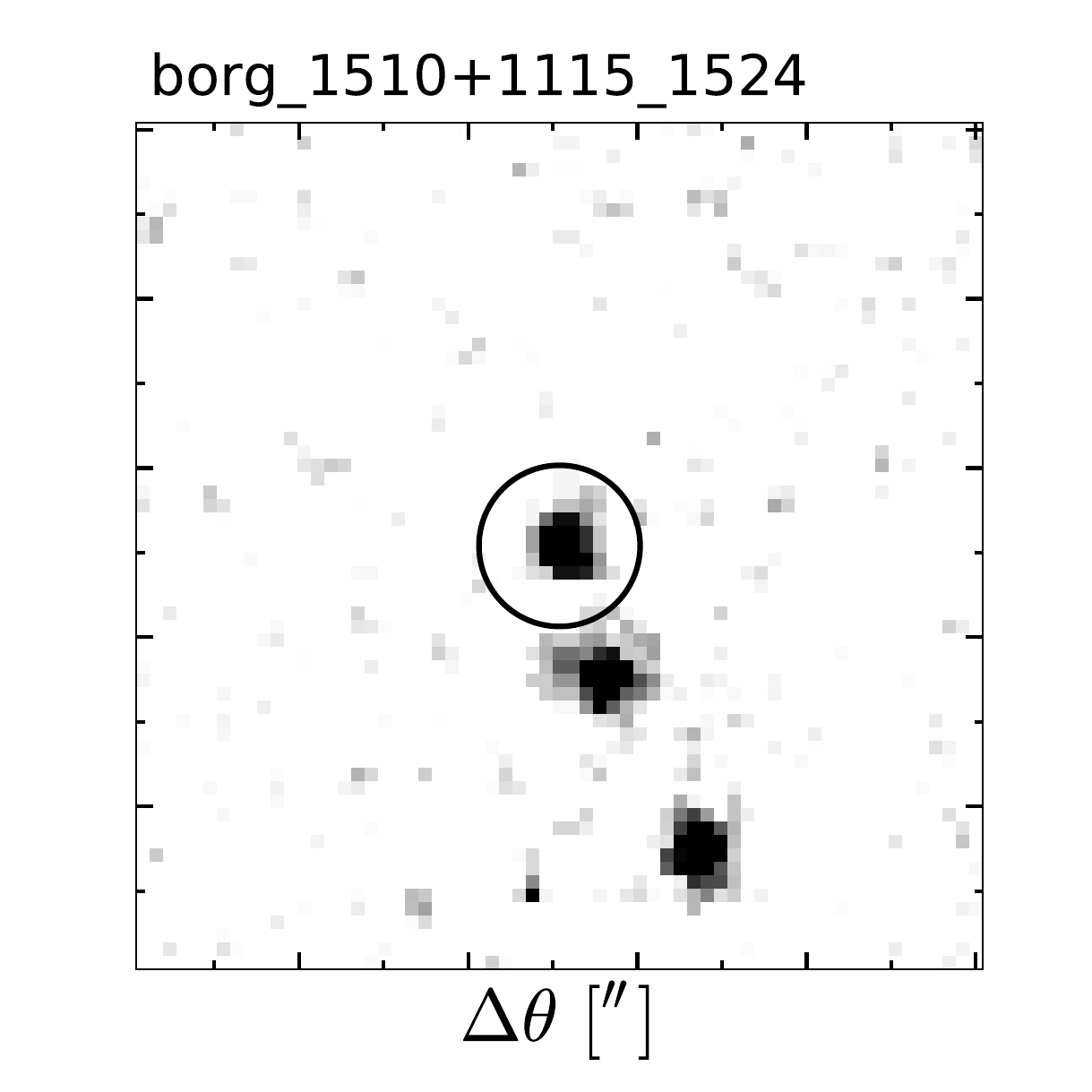}
\includegraphics[trim=0 15 50 0, clip, scale=.175]{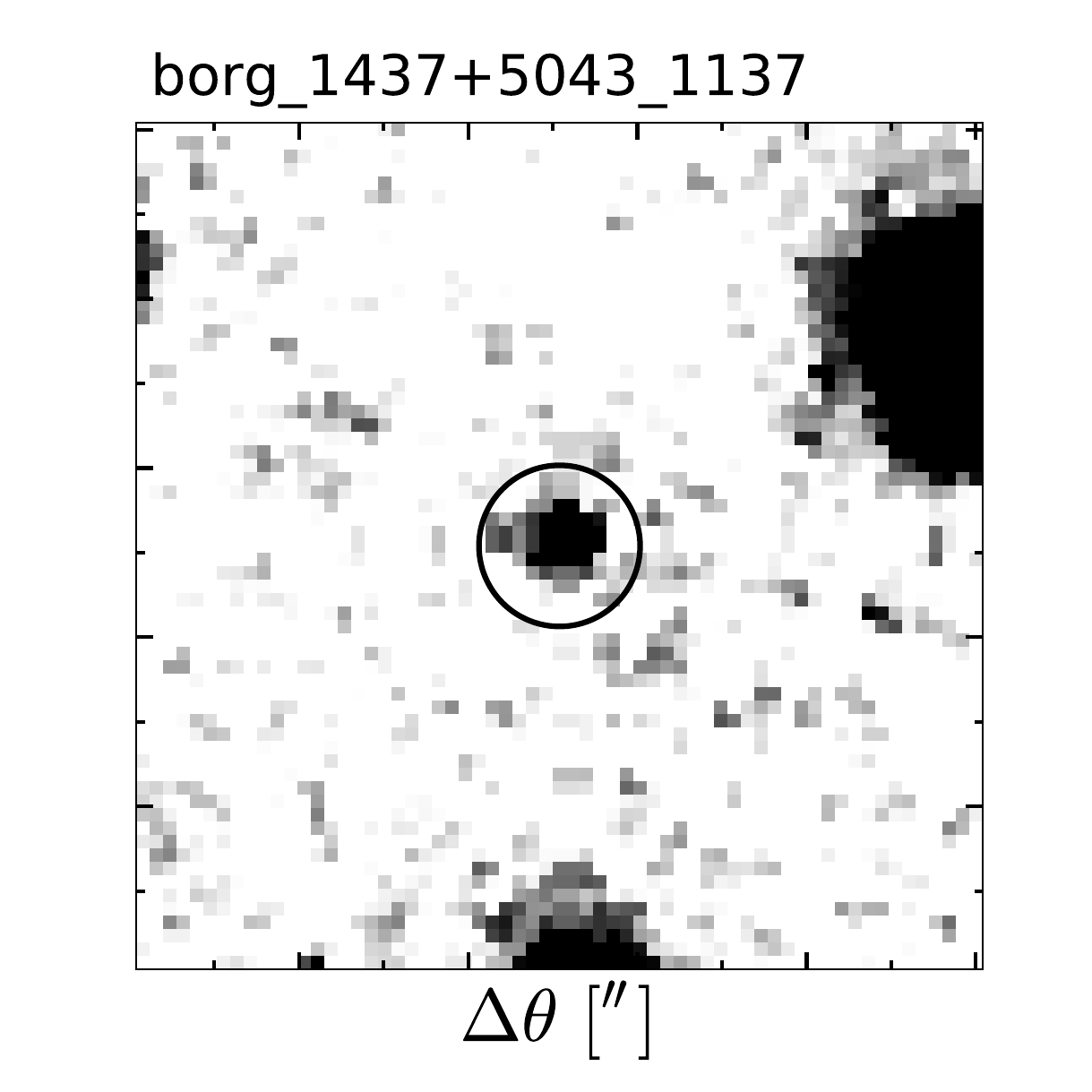}
\includegraphics[trim=0 15 50 0, clip, scale=.175]{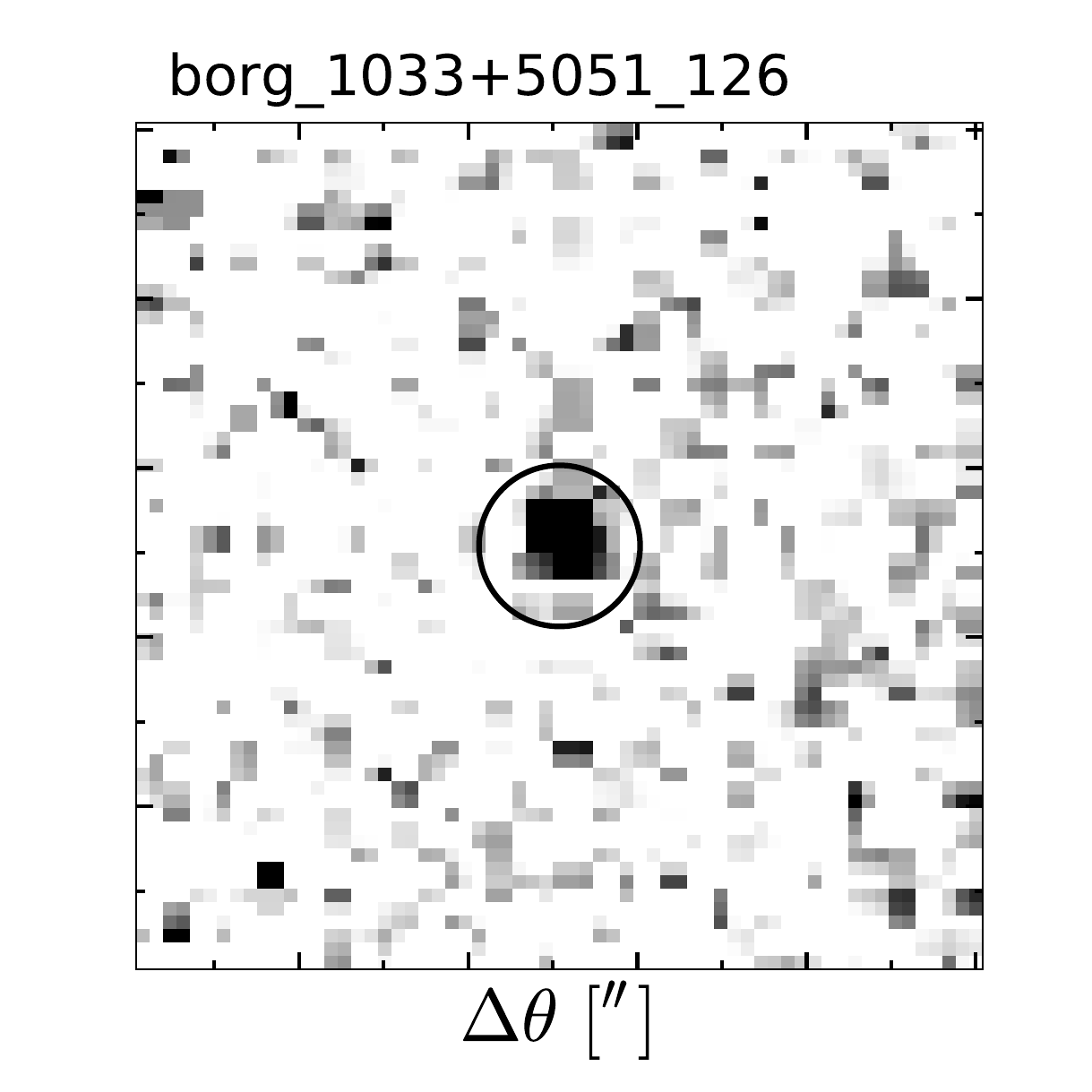}
\caption{The 10 $8\sigma$ $z\sim8$ LBG candidates identified in the BoRG survey. The cutouts are $5$ arcsec on each side, which is the characteristic scale for galaxy-galaxy lensing \citep{barone2015impact}, and shown in the $J_{125}$ band. Seven of the 10 candidates are ruled out as being strongly gravitationally lensed due to the absence of foreground objects along the line of sight. The object borg$\_1301$+$0000\_160$ appears nearby a bright foreground galaxy, and \citet{mason2015correcting} assess its magnification to be $\mu=1.47\pm0.30$. The object borg$\_1437$+$5043\_1137$ appears $2.6$ arcsec from a bright foreground at $z_{phot}=1.4$. Using the framework of \citet{barone2015impact}, we assess this object's magnification to be $\mu=1.3\pm0.2$ and for it to have a $5$ per cent chance of being strongly lensed. Neither of these configurations display obvious secondary images, which is the key criterion for constraining strong lensing. The object borg$\_0440$-$5244\_682$ is in close proximity to a group of three foreground galaxies, around which a candidate secondary image is identified.}
\label{figure:allcutouts}
\end{figure*}

\begin{figure*}
\includegraphics[trim=0 0 80 0, clip, scale=0.115]{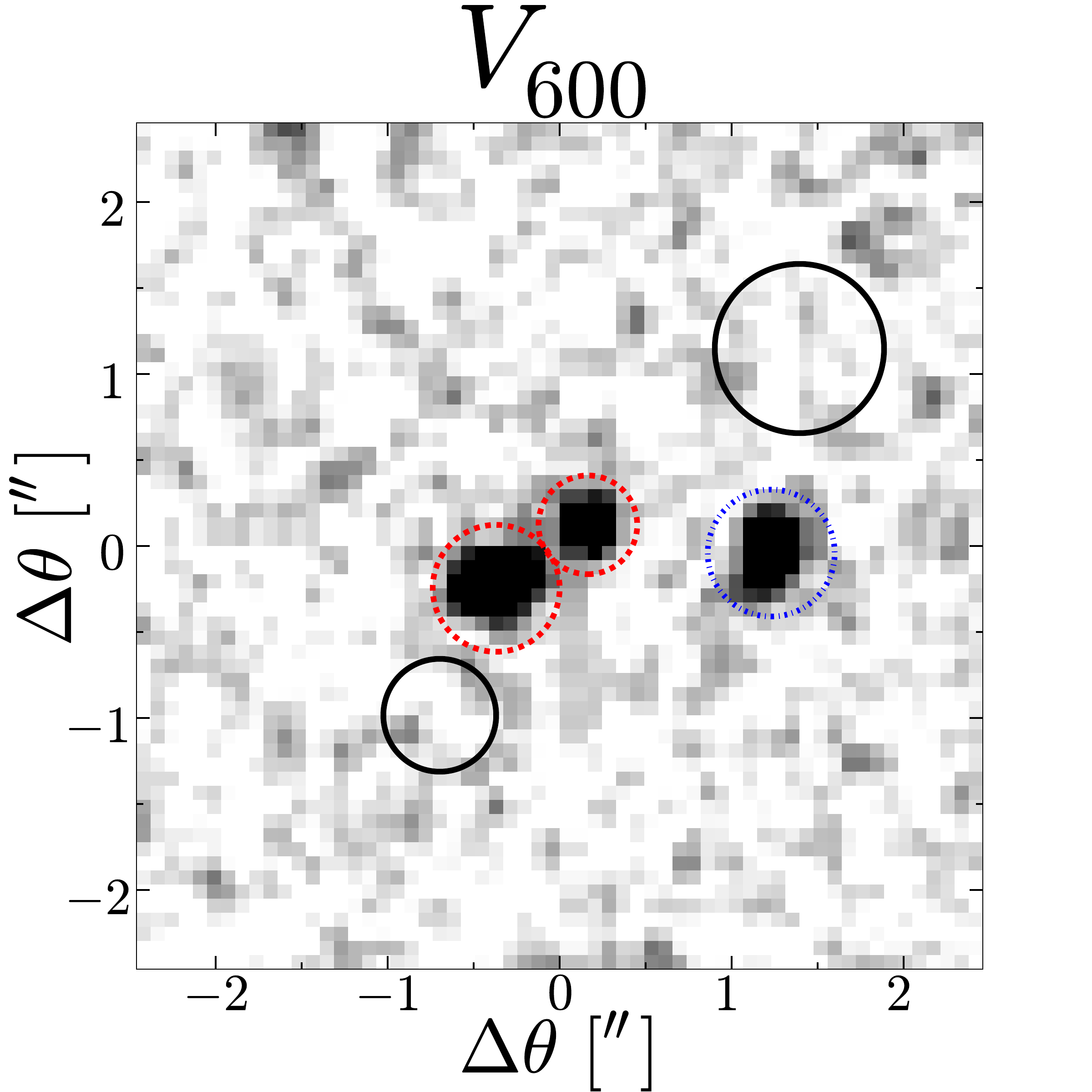}
\includegraphics[trim=10 0 80 0, clip, scale=0.115]{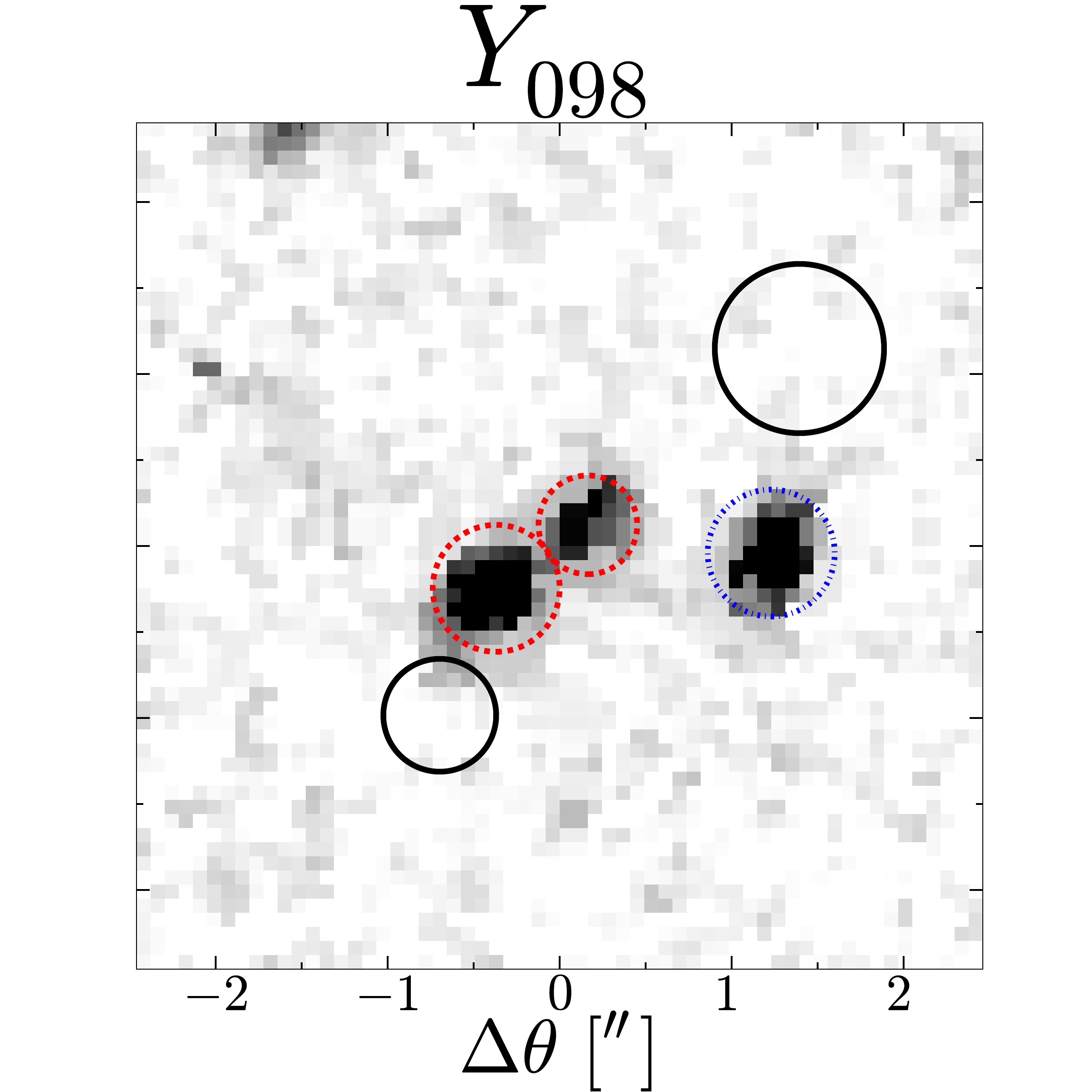}
\includegraphics[trim=10 0 80 0, clip, scale=0.115]{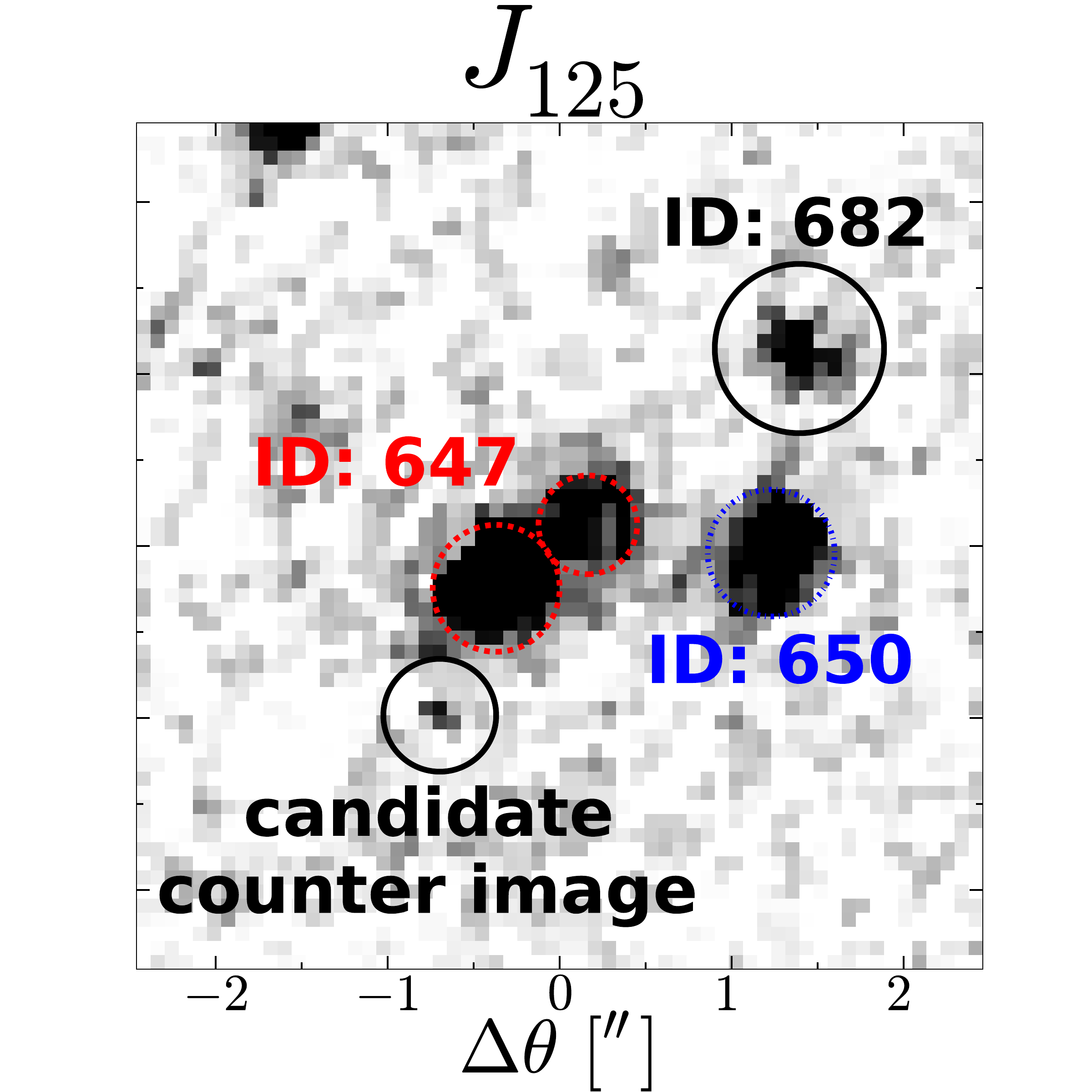}
\includegraphics[trim=10 0 65 0, clip, scale=0.115]{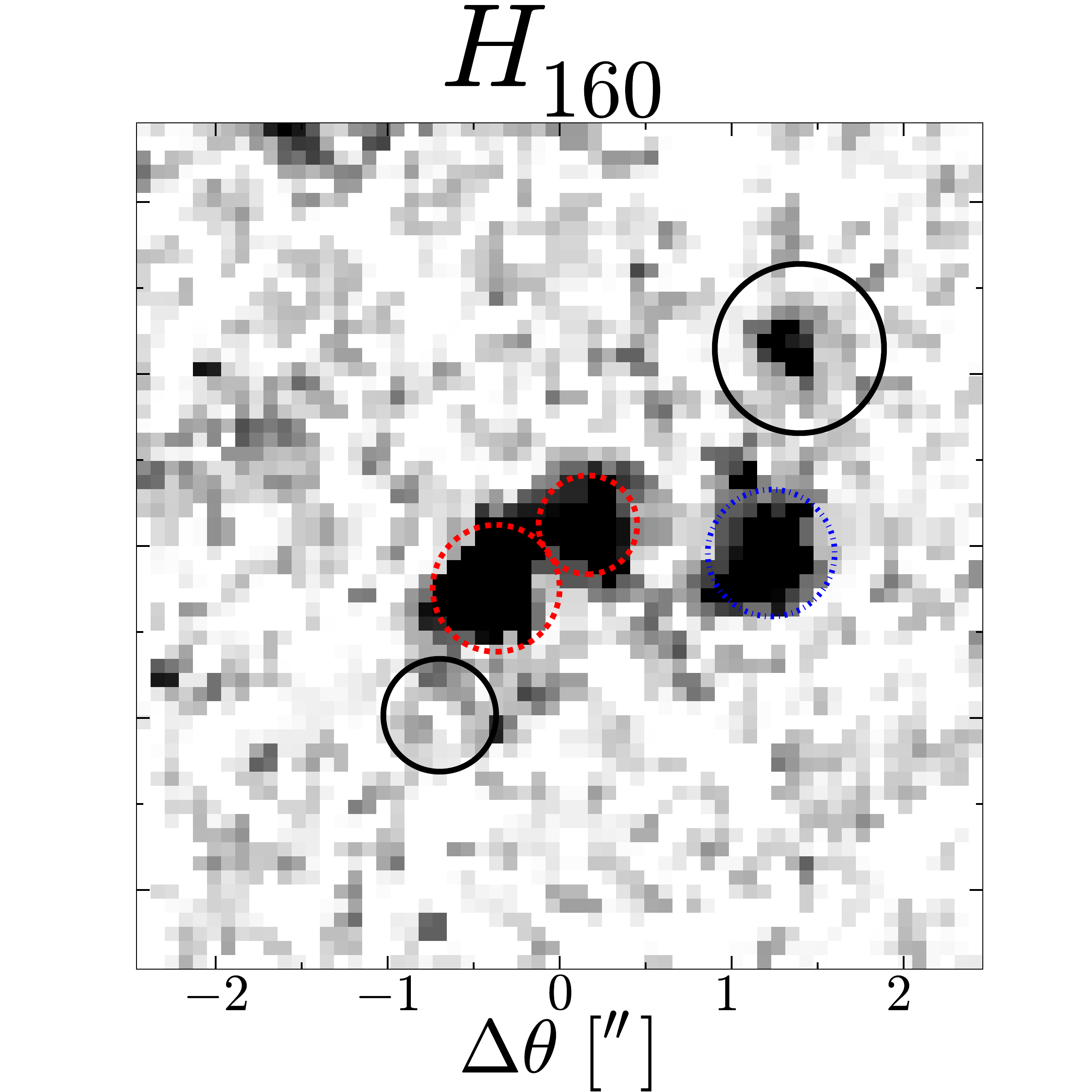}
\caption{Postage stamp images of the candidate lensed $z\sim8$ galaxy in field BoRG$\_0440$-$5244$ in the four HST/WFC3 filters (left to right: $V_{600}$, $Y_{098}$, $J_{125}$, $H_{160}$). The two central foreground galaxies (ID: 647) are circled in red (dashed) and the third foreground galaxy (ID: 650) is circled in blue (dash-dotted). The location of the $Y_{098}$-dropout galaxy is marked by the larger black circle (solid), and the possible faint counter image on the opposite side of the lens is indicated by the smaller black circle (solid). The images are $5.0$ arcsec on each side. All cutouts are shown on the same contrast scale.}
\label{figure:filter_comp}
\end{figure*}

\indent The large number of independent fields in BoRG results in a sample essentially unaffected by large-scale structure uncertainty \citep{trenti2008cosmic}, and includes a variety of different environments in terms of line-of-sight structure, representing an ideal dataset to quantify magnification bias in a different manner to that used by \citet{barone2015impact} and \citet{mason2015correcting}. To search for evidence of large magnifications, we have visually inspected the regions surrounding the 10 sources detected with a S/N $\geq 8\sigma$ in the BoRG survey, all of which have $m_{J_{125}}<27.0$. We investigate the highest S/N objects because they offer the best chance of identifying direct evidence of strong lensing in the event that they have been significantly magnified. The indications of strong gravitational lensing that we inspect for are proximity of the dropout to bright foreground objects, and secondary images with similar colours to the dropout. Cutouts of the 10 candidates in the $J_{125}$ band are shown in Fig. \ref{figure:allcutouts}.\\
\indent We identify one good candidate for being a strongly gravitationally lensed LBG at $z\sim8$ (borg$\_0440$-$5244\_682$ in the notation of \citealt{bradley2012brightest}), which appears very close in projection to a compact grouping of three foreground galaxies (borg$\_0440$-$5244\_647$, which consists of two objects, and borg$\_0440$-$5244\_650$). Through visual inspection of the configuration, we identify a faint (S/N $\sim2$) possible counter image, which appears in the deepest band, $J_{125}$, and is located immediately below the foreground galaxies opposite the location of the dropout. The location and flux relative to the brighter image is consistent with gravitational lensing theory, which predicts secondary images to appear demagnified on the opposite side of the deflector, and at a smaller separation from the deflector than the brighter image. In fact, a strikingly similar gravitational lensing configuration has been observed \citep[][see fig. 3]{wong2014discovery} with almost identical relative image positions, magnifications and flux ratio, albeit with a lower redshift source and smaller deflection angles.\\
\indent The proximity to multiple bright foreground galaxies and identification of a possible secondary image with similar colours to the dropout makes borg$\_0440$-$5244\_682$ a good candidate to be magnified significantly, warranting further investigation. Postage stamp images of the dropout and foreground galaxies in the single optical and three near-IR filters of the BoRG survey are presented in Fig. \ref{figure:filter_comp}, where the objects are labelled in the $J_{125}$ panel.\\
%

\subsection{Photometry of the dropout and foreground objects}
\label{subsection:photometry}
\indent The candidate $z\sim8$ galaxy, borg$\_0440$-$5244\_682$, was identified in the BoRG field borg$\_0440$-$5244$. Identification of $z\sim8$ dropouts is made according to strict colour selection. In general, $z\gtrsim7.5$ galaxies are identified due to a strong break between the $Y_{098}$ and $J_{125}$ filters, a blue or flat $J_{125}-H_{160}$ colour and a non-detection in the optical $V_{600}$ band. The selection criteria requires,
\begin{eqnarray}
&S/N_{V_{600}} < 1.5\nonumber\\
&(Y_{098} - J_{125}) > 1.75\nonumber\\
&(J_{125} - H_{160}) < 0.02 + 0.15[(Y_{098} - J_{125}) - 1.75]\nonumber.
\end{eqnarray}
\indent The conservative non-detection limit of $1.5\sigma$ in the $V_{600}$ filter is crucial in producing a clean $z\sim8$ sample \citep{bouwens2011ultraviolet}. The photometry presented by \citet{bradley2012brightest} of the candidates analyzed here gives a $J_{125}$ magnitude of $25.9\pm0.1$ mag, a $Y_{098}-J_{125}$ residual of $>2.1$ mag, and S/N after sky subtraction in $V_{600}$, $Y_{098}$, $ J_{125}$ and $H_{160}$ of $-0.8, -0.5, 9.1$ and $5.7$, respectively. We find a $J_{125}$ ellipticity for the dropout of $\epsilon=0.4\pm0.1$ tangential to the direction of the largest of the proposed lensing galaxies. The $J_{125}$ half-light radius of the dropout was measured, using $\tt{SExtractor}$, to be $0.21$ arcsec ($1.1$kpc at $z\sim8$; not corrected for point spread function (PSF) broadening).\\
\indent The two brighter foreground objects are identified as a single blended source (borg$\_0440$-$5244\_647$) with $m_{J_{125}}=23.41\pm0.02$ mag in the BoRG SExtractor catalogue. We performed fixed aperture photometry on the individual galaxies of borg$\_0440$-$5244\_647$ and found the brighter galaxy to be $1.3$--$1.5$ mag brighter than the fainter galaxy in each of the filters, making the colours of the two galaxies essentially identical. The galaxy borg$\_0440$-$5244\_650$, has $m_{J_{125}}\sim24.58\pm0.14$ mag. The photometry of all objects is summarised in Table \ref{table:photometry}. \\

\subsection{Photometry of the candidate counter image}
\label{subsection:counterphot}
\begin{figure}
\begin{center}
\includegraphics[trim=0 0 80 0, clip, scale=0.110]{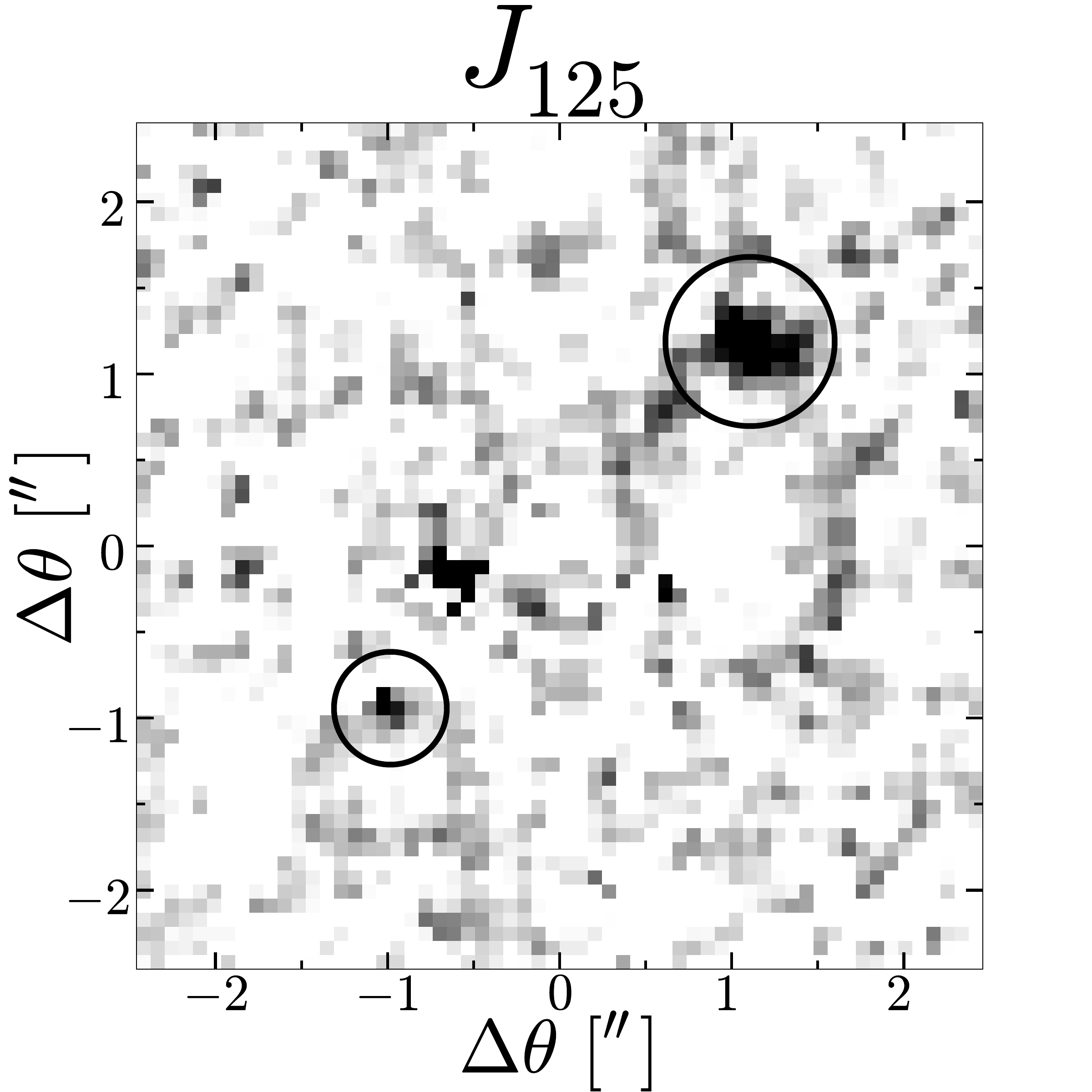}
\includegraphics[trim=80 0 0 0, clip, scale=0.110]{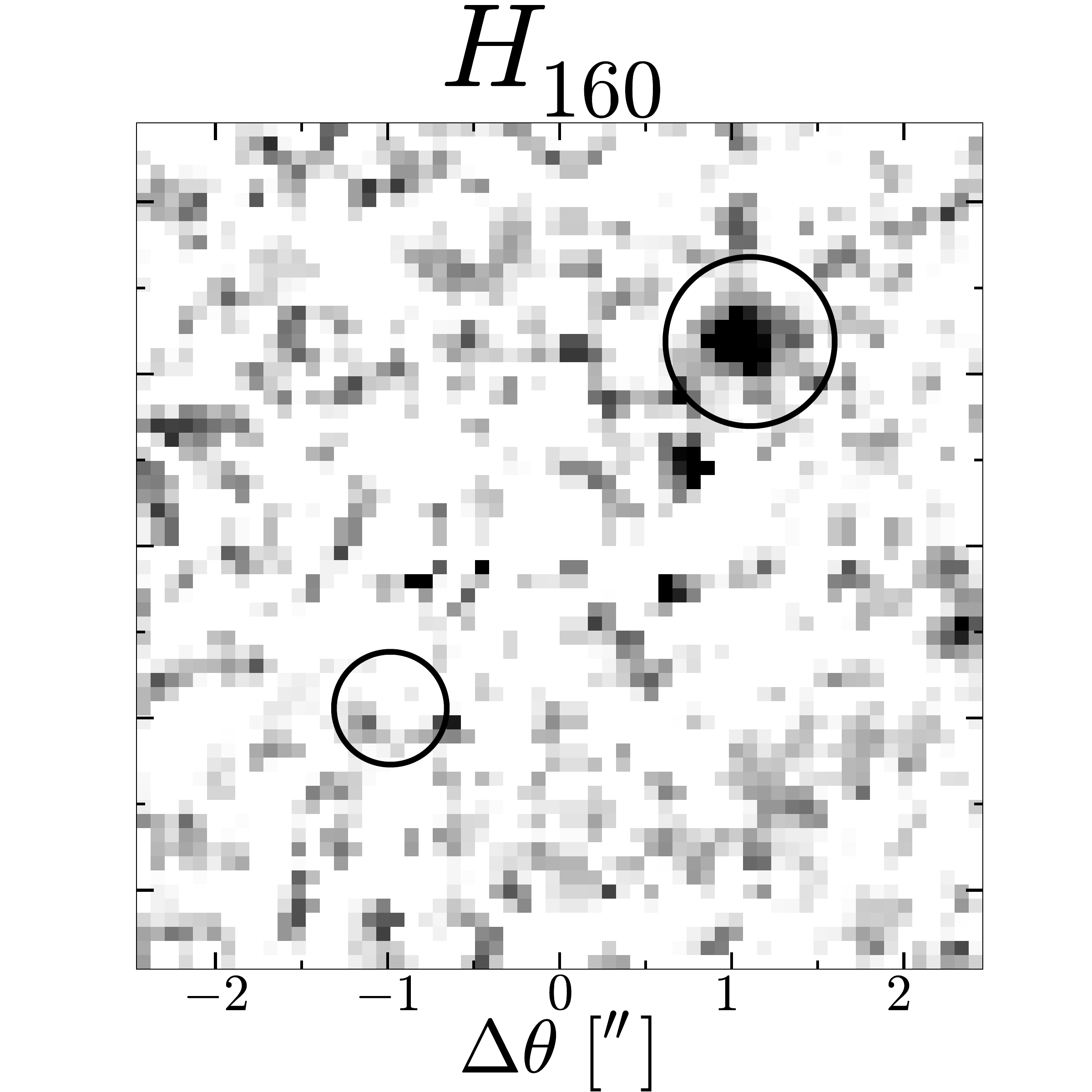}
\caption{Normalized residuals for the subtractions of the PSF-matched $Y_{098}$ image from the $J_{125}$ image (left) and $H_{160}$ image (right). The images are $5.0$ arcsec on each side. There remains some residual of the deflector galaxy cores due to the varying light profiles of the galaxies between filters. There is a marginal detection of the counter image (S/N$\sim2$) in $J_{125}$. As expected, the counter image is undetected in the $H_{160}$ filter. The primary and candidate counter images are both circled in black.}
\label{figure:sub_comp}
\end{center}
\end{figure}

\indent In addition to the photometry already carried out by \citet{bradley2012brightest}, we perform aperture photometry in $\tt{IRAF}$ on the primary and counter images to determine their flux ratio and the counter image's $Y_{098}-J_{125}$ and $V_{600}-J_{125}$ colours. To minimize contamination of the counter image photometry by the lensing group, we operate on a version of the $J_{125}$ and $H_{160}$ images where we subtracted a scaled and PSF-matched version of the $Y_{098}$ data, where the $z\sim8$ candidate has dropped out. For this, we use $\tt{ISIS2.2}$ \citep{alard2000image} to produce a convolution kernel between the images in the different bands and to match the PSFs. There is some residual in the subtracted images due to the foreground galaxies having intrinsically different profiles between filters. The galaxies on the whole, though, are adequately subtracted, and the residuals do not affect the photometry at the image positions. The subtracted images are shown in Fig. \ref{figure:sub_comp}.\\
\indent Photometry is performed in a $4$ pixel circular aperture ($0.32$ arcsec). In $J_{125}$, the integrated flux in the primary image is $1.54\pm0.18\times10^{-30}$ erg cm$^{-2}$ s$^{-1}$ Hz$^{-1}$ (consistent with the S/N$=9.1$ determination in \citealt{bradley2012brightest}), and the flux in the counter image is $3.21\pm1.72\times10^{-31}$ erg cm$^{-2}$ s$^{-1}$ Hz$^{-1}$. The uncertainty includes both Poisson noise and background. This gives a flux ratio of $4.8\pm2.6$, where the large error propagates from the low S/N of the counter image.\\
\indent We measure a $Y_{098}-J_{125}$ colour redder than $1.2$ mag and a $V_{600}-J_{125}$ colour redder than $0.5$ mag for the counter image. When the $Y_{098}$ and $V_{600}$ images are stacked, the $(Y_{098}+V_{600})-J_{125}$ colour is redder than $1.4$ mag. While BoRG is not designed to detect such faint objects, and neither of these colours satisfy the formal BoRG criteria for $z\sim8$ LBG selection, the colours are consistent with both the brighter image, and the source being at $z\sim8$ but beyond the $5\sigma$ limiting flux of this field.\\
\indent We do not detect a signal at the location of the counter image in the $H_{160}$ image, however this is not surprising as the flux ratio between the primary and counter image is measured in $J_{125}$ to be $4.8\pm2.6$, and the $H_{160}$ primary image is detected at $5.7\sigma$, hence we only expect the $H_{160}$ counter image to be undetected (S/N $=1.2\pm0.7$). Inspection of the individual flat-fielded files in $J_{125}$ does not suggest that the faint candidate counter image is a detector artifact.\\
\indent The non-detection in $H_{160}$ is taken into account in our Markov-Chain Monte Carlo analysis (Section \ref{section:MCMC}), as both bands are fitted simultaneously.\\
%
\begin{table}
\scalebox{0.94}{
\begin{tabular}{|lcccc|}
\hline
ID 	&$V_{600}$&$Y_{098}$& $J_{125}$&$H_{160}$\\
\hline
\hline
682&$>28.26$&$>28.15$&$25.92\pm0.14$& $25.75\pm0.19$\\	
647&$24.20\pm0.07$&$23.87\pm0.03$&$23.41\pm0.02$&$23.49\pm0.04$\\
650&$25.17\pm0.13$&$24.87\pm0.05$& $24.58\pm0.14$&$24.31\pm0.31$\\
\hline
\end{tabular}}
\caption{The photometry for the primary dropout (ID: 682), the main lensing group (ID: 647) and a nearby galaxy (ID: 650) which is potentially part of the group. The main lensing group (647) has a spectroscopically confirmed redshift of $z=1.327$. The photometric redshift probability distribution of the third object (650) is extended, however it has a non-negligible probability of residing at the same redshift as the central galaxies.}
\label{table:photometry}
\end{table}

\subsection{Spectroscopy of the central foreground galaxies and dropout}
\label{subsection:spectroscopy}
\indent The source and foreground galaxy configuration and image flux ratios outlined in the previous sections and Fig. \ref{figure:filter_comp} present the exciting possibility that this system is a $z\sim8$ galaxy gravitationally lensed by a foreground galaxy group. In order to further strengthen this interpretation, we obtained spectroscopic observations of the foreground galaxy group and $z\sim8$ LBG candidate. The goals of the spectroscopic measurements of the system are two-fold; The first goal was to obtain spectroscopic redshifts for the two foreground galaxies (ID: 647), and the second goal was to either identify emission lines, such as Lyman-alpha emission, at the location of the $z\sim8$ LBG candidate (ID: 682), or use the lack of emission lines to constrain its redshift in conjunction with its photometric colours \citep[e.g.][]{treu2013changing}.\\
\indent In order to obtain these measurements, we carried out VLT/X-Shooter spectroscopic observations in visitor mode on August 13-15, 2013 (program \#091.A-0053, PI Puzia). Observing conditions were excellent with seeing in the range $0.6$--$0.9$ arcsec. The X-Shooter long slit ($11$ arcsec) was positioned to cover both the central pair of foreground deflectors and the Y-dropout. X-Shooter provides an uninterrupted wavelength coverage in the range $300~\mathrm{nm}< \lambda < 2480~\mathrm{nm}$, allowing us to characterise emission lines both from the deflector and the dropouts with a single observation. A total exposure time of 6 hours (3 hours on target) was achieved on the source. Telluric standard stars were observed for calibration purposes throughout the duration of the observations. The data was then reduced using the European Southern Observatory (ESO) X-Shooter pipeline v.2.5.2 \citep{modigliani2010x} and the Gasgano data file organizer developed by ESO.\\

\subsubsection{Emission lines from the foreground galaxies}
\begin{figure*}
\begin{center}
\includegraphics[trim=4 0 4 0, clip, scale=1.2]{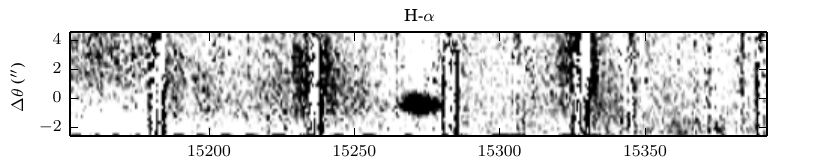}\\
\includegraphics[trim=4 0 4 0, clip, scale=1.2]{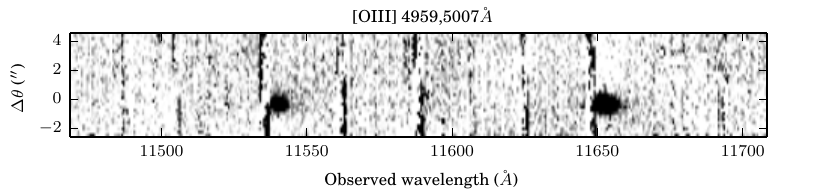}
\caption{The cropped two-dimensional spectrum of the central deflector galaxies, obtained using X-Shooter on the VLT. Top: The hydrogen-alpha emission line, indentified at $\lambda=15272\textrm{\AA}$. Bottom: The [OIII] doublet emission lines, identified at $\lambda=11540\textrm{\AA}$ and $\lambda=11651\textrm{\AA}$. These emission lines provide a robust spectroscopic redshift of $z=1.327$ for the deflector galaxies.}
\label{figure:emlines}
\end{center}
\end{figure*}

\indent We search for emission lines at the location of the two central foreground galaxies to determine their redshifts. The two galaxies have a separation of $0.7$ arcsec, and the seeing of the observations was $0.6$--$0.9$ arcsec, therefore the two galaxies are unresolved in the spectrum. We identify the redshifted hydrogen-alpha $6563\textrm{\AA}$, hydrogen-beta $4861\textrm{\AA}$, [OIII] $5007\textrm{\AA}$ and $4959\textrm{\AA}$ doublet and the [OII] $3726\textrm{\AA}$ and $3729\textrm{\AA}$ doublet emission lines at the location of the two foreground galaxies. The two-dimensional (2D) spectrum of the hydrogen-alpha and [OIII] doublet emission lines are shown in Fig. \ref{figure:emlines}. The observed emission lines unambiguously places the foreground galaxy pair at $z=1.327$, which supports the strong gravitational lensing hypothesis. This redshift is at the peak of the redshift distribution for the deflectors of $z\sim7$-$8$ LBGs \citep{barone2015impact}. \\
\indent It should be noted that there is a chance that the emission lines are coming only from the brighter foreground galaxy. However, the colours of the two central galaxies are virtually identical and no emission lines are observed at the location of the fainter galaxy elsewhere in the spectrum, leading to the likely possibility that the two galaxies are indeed a close pair at $z=1.327$.\\

\subsubsection{Constraints on the redshifts of the LBG}
\label{subsubsection:dropoutspec}
\begin{figure*}
\begin{center}
\includegraphics[trim=0 0 0 0, clip, scale=0.7]{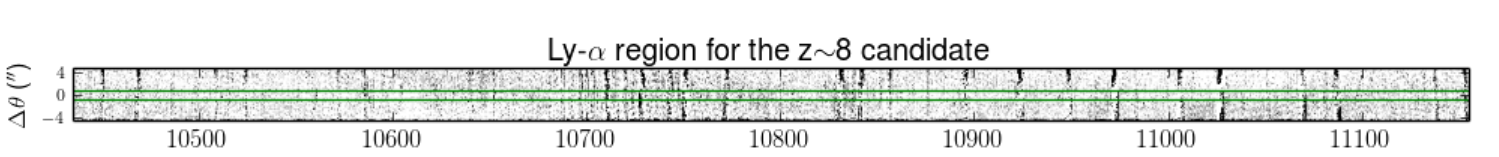}\\
\includegraphics[trim=0 0 0 0, clip, scale=0.7]{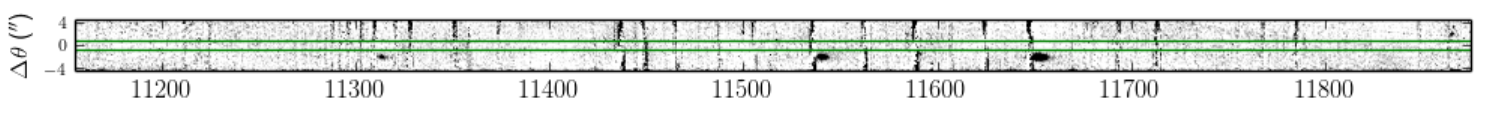}\\
\includegraphics[trim=0 0 0 0, clip, scale=0.7]{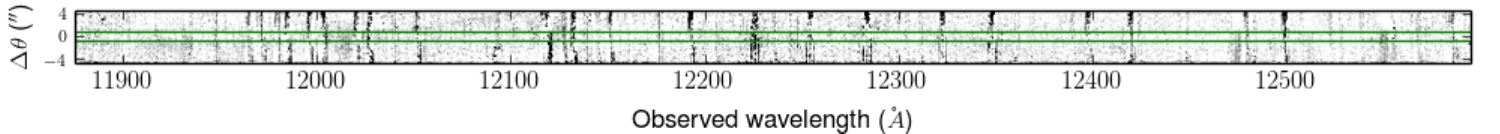}\\
\vspace{5mm}
\includegraphics[trim=0 0 0 0, clip, scale=0.7]{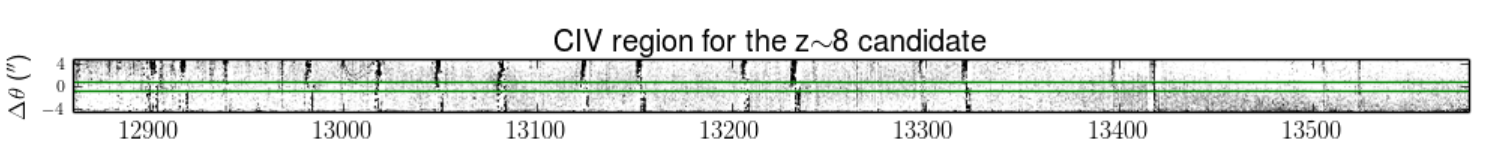}\\
\includegraphics[trim=0 0 0 0, clip, scale=0.7]{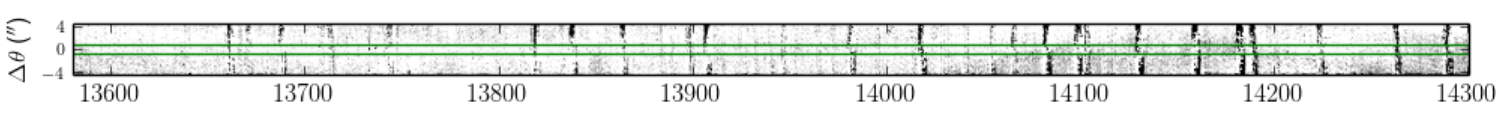}\\
\includegraphics[trim=0 0 0 0, clip, scale=0.7]{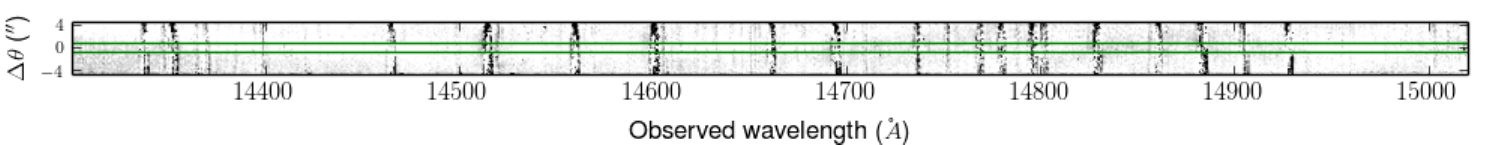}\\
\vspace{5mm}
\includegraphics[trim=0 0 0 0, clip, scale=0.7]{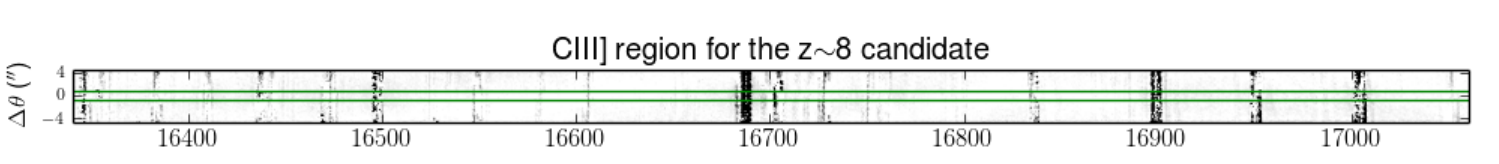}\\
\includegraphics[trim=0 0 0 0, clip, scale=0.7]{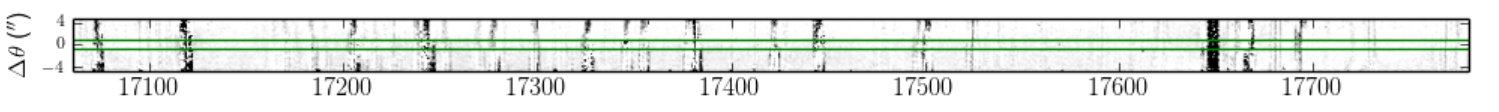}\\
\includegraphics[trim=0 0 0 0, clip, scale=0.7]{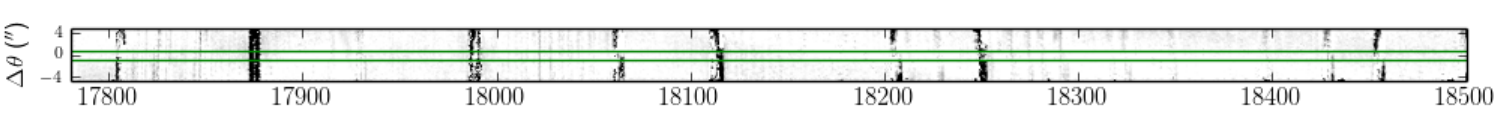}\\
\includegraphics[trim=0 0 0 0, clip, scale=0.7]{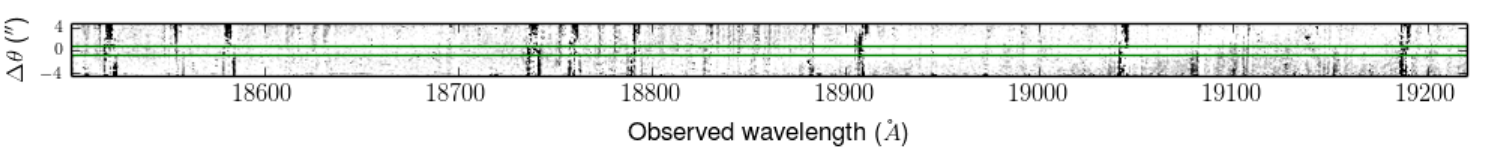}\\
\caption{The regions of the 2D spectrum where the Lyman-alpha ($1216\textrm{\AA}$, top), CIV ($1548\textrm{\AA}$, middle), and CIII] ($1909\textrm{\AA}$, bottom) emission lines could appear given the redshift distribution of the dropout. The portion of the spectrum covering the dropout falls within the green lines. No lines are detected at these wavelengths or elsewhere in the spectrum at the location of the dropout. We use these non-detections to rule out a major class of low-redshift interlopers, and constrain the equivalent widths of the Lyman-alpha, CIII] and CIV emission lines. The hydrogen-beta and [OIII] doublet emission lines from the central galaxy pair are visible in the Lyman-alpha detection region at $11312\textrm{\AA}$, $11540\textrm{\AA}$ and $11651\textrm{\AA}$, respectively.}
\label{figure:z8spec}
\end{center}
\end{figure*}

\indent Spectroscopic confirmation of high redshift galaxies has been very difficult in the past, hence using gravitationally lensed LBGs provides the best opportunity to probe intrinsically fainter sources at high redshift. The $z\sim8$ candidate LBG was included in our spectrum in order to search for the Lyman-alpha emission line. The X-Shooter wavelength coverage would allow detection of the Lyman-alpha line, from a Lyman-alpha emitter at $z\gtrsim7$, as well as other emission lines such as CIII] and CIV. However we do not detect the Lyman-alpha line in the spectrum, nor do we detect any other emission lines. For an LBG residing at $7.6\lesssim z \lesssim 8.8$ (the redshift where there is a non-negligible probability density for this LBG), we show the 2D spectrum of the $Y_{098}$ dropout at the wavelengths where Lyman-alpha, CIII] and CIV could conceivably appear in Fig. \ref{figure:z8spec}. We place upper limits on the equivalent widths of these lines in Section \ref{subsubsection:ews}.\\
\indent Despite the lack of visible emission lines, the spectrum of the LBG candidate is nevertheless useful for constraining the strong lensing hypothesis since we can rule out the possibility of the source being a strong emission line contaminant. We do not observe any emission lines at the location of the dropout down to the sensitivity limit of the our spectroscopic observations. While this prevents us from achieving spectroscopic confirmation, the non-detection of emission lines is important to exclude the most significant class of contaminants in samples of high redshift LBGs. These contaminants are objects with a very weak continuum and very strong hydrogen-alpha, hydrogen-beta and [OIII] emission located at $1.3\lesssim z\lesssim1.7$. Such galaxies could exhibit colours similar to those observed in the $z\sim8$ candidate of this paper \citep{atek2011very}.\\ 
\indent We consider the noise levels of the dropout spectrum to place limits on the contribution to the observed broadband flux from emission lines. The integrated line flux at $\lambda\approx 1.6~\mathrm{\mu m}$ required to produce a broadband flux of $m_{H_{160}}=25.75$ mag is $7.8\times10^{-17}$ erg cm$^{-2}$ s$^{-1}$. Assuming a simple model of an emission line with a width of $10\textrm{\AA}$, the required flux density to account for all of the broadband magnitude is $7.8\times10^{-18}$ erg cm$^{-2}$ s$^{-1}$ $\textrm{\AA}^{-1}$. The median noise in the spectrum in the $H_{160}$ band is $4.6\times10^{-19}$ erg cm$^{-2}$ s$^{-1}$ $\textrm{\AA}^{-1}$. This allows us to rule out the presence of an emission line contributing more than $m_{H_{160}}=28.6$ at $5\sigma$, and contributing all of the observed broadband flux of $m_{H_{160}}=25.75$ mag at $70\sigma$. Similarly, the integrated line flux required to produce a broadband flux of $m_{J_{125}}=25.92$ mag is $7.9\times10^{-17}$ erg cm$^{-2}$ s$^{-1}$, and the median noise level in the spectrum covering the $J_{125}$ band is $1.0\times10^{-18}$ erg cm$^{-2}$ s$^{-1}$ $\textrm{\AA}^{-1}$. This allows us to rule out the presence of an emission line contributing more than $m_{J_{125}}=27.9$ mag at $5\sigma$, and contributing the all of the observed broadband flux of $m_{J_{125}}=25.92$ mag at $30\sigma$.\\
\indent We calculate the likelihood that the emission lines are hidden behind sky lines or obscured by atmospheric absorption. We define the obscured part of the spectrum as the region where an emission line detected at $5\sigma$ at the median noise would be detected at less than $2\sigma$. The percentage of the spectrum covering the broadband $J_{125}$ and $H_{160}$ filters obscured by sky lines or atmospheric absorption above the required level of emission lines is $\sim14$ per cent. If the dropout resides at $1.3\lesssim z\lesssim1.7$, the likelihood that both the brighter [OIII] emission line and hydrogen-alpha line are obscured is $10$ per cent. The likelihood that the hydrogen-alpha line and both of the [OIII] lines are obscured is $7$ per cent. \\ 
\indent We conclude through this simple analysis that emission lines from the dropout source can contribute at most $\sim20$ per cent of the broadband flux in $J_{125}$ and less than $\sim10$ per cent in $H_{160}$, therefore excluding lower redshift solutions with a faint (undetected) continuum and broadband photometry dominated by strong emission lines with very high confidence.\\
\indent As a consistency check, we compare the required luminosity of the emission lines in the case of the dropout being a low redshift interloper with the detected emission lines of the foreground galaxies. For emission lines to contribute the required flux in $J_{125}$ and $H_{160}$ for the dropout, they would need to be as bright as the observed [OIII] $4959\textrm{\AA}$ line from the foreground galaxy pair, and twice as bright as the observed hydrogen-beta line, both of which are detected with high confidence ($\gtrsim 10\sigma$).\\

\subsubsection{Equivalent width constraints from the LBG spectrum}
\label{subsubsection:ews}
\indent In addition to ruling out strong emission line interlopers, the absence of emission lines allows us to place upper limits on the equivalent widths of the Lyman-alpha, CIII] and CIV emission lines for this LBG. Each of these emission lines have been observed in spectra of LBGs at $z\gtrsim6$ previously \citep{ono2011spectroscopic, finkelstein2013galaxy, oesch2015spectroscopic, stark2015spectroscopic, stark2015spectroscopic2}.\\ 
\indent We investigate the limits that can be placed on the equivalent width of Lyman-alpha emission. We do this by assuming a half-Gaussian model of the Lyman-alpha emission line profile with a full width at half maximum (FWHM) of $12\textrm{\AA}$. We calculate the $5\sigma$ upper limit on the integrated line flux by considering the noise in the spectrum in the possible wavelength range of Lyman-alpha emission. The continuum flux density is computed from the broadband photometry. The result is a median $5\sigma$ upper limit of the rest-frame equivalent width of EW$_0($Lyman-$\alpha)<57\textrm{\AA}$. We quantify the chance that Lyman-alpha is  obscured by sky lines or atmospheric absorption in the region where we expect Lyman-alpha to appear. The redshift distribution for the LBG is centred on $z=8$, but there is a non-negligible probability density from $z\sim7.6$ to $z\sim8.8$. Using the range of wavelengths at which Lyman-alpha could appear, we calculate that the chance that it has been obscured to be $14$ per cent. \\
\indent We also constrain the equivalent width of CIII] and CIV emission lines. For these emission lines, we assume a Gaussian emission line profile model with a FWHM of $12\textrm{\AA}$. We calculate the $5\sigma$ upper limit on the integrated line flux by considering the noise in the spectrum in the possible wavelength range of CIII] and CIV emission. The continuum flux density is again calculated from the broadband photometry. The results are median $5\sigma$ upper limits for CIII] and CIV emission of EW$_0($CIII]$)<25\textrm{\AA}$ and EW$_0($CIV$)<36\textrm{\AA}$, respectively. We quantify the chance of the CIII] and CIV lines being obscured by sky lines or due to atmospheric absorption as $12$ per cent in both cases.\\

\subsubsection{Photometric redshift of the third foreground galaxy}
\label{subsubsection:photoz}
\begin{figure}
\begin{center}
\includegraphics[trim=0 0 0 0, clip, scale=0.4]{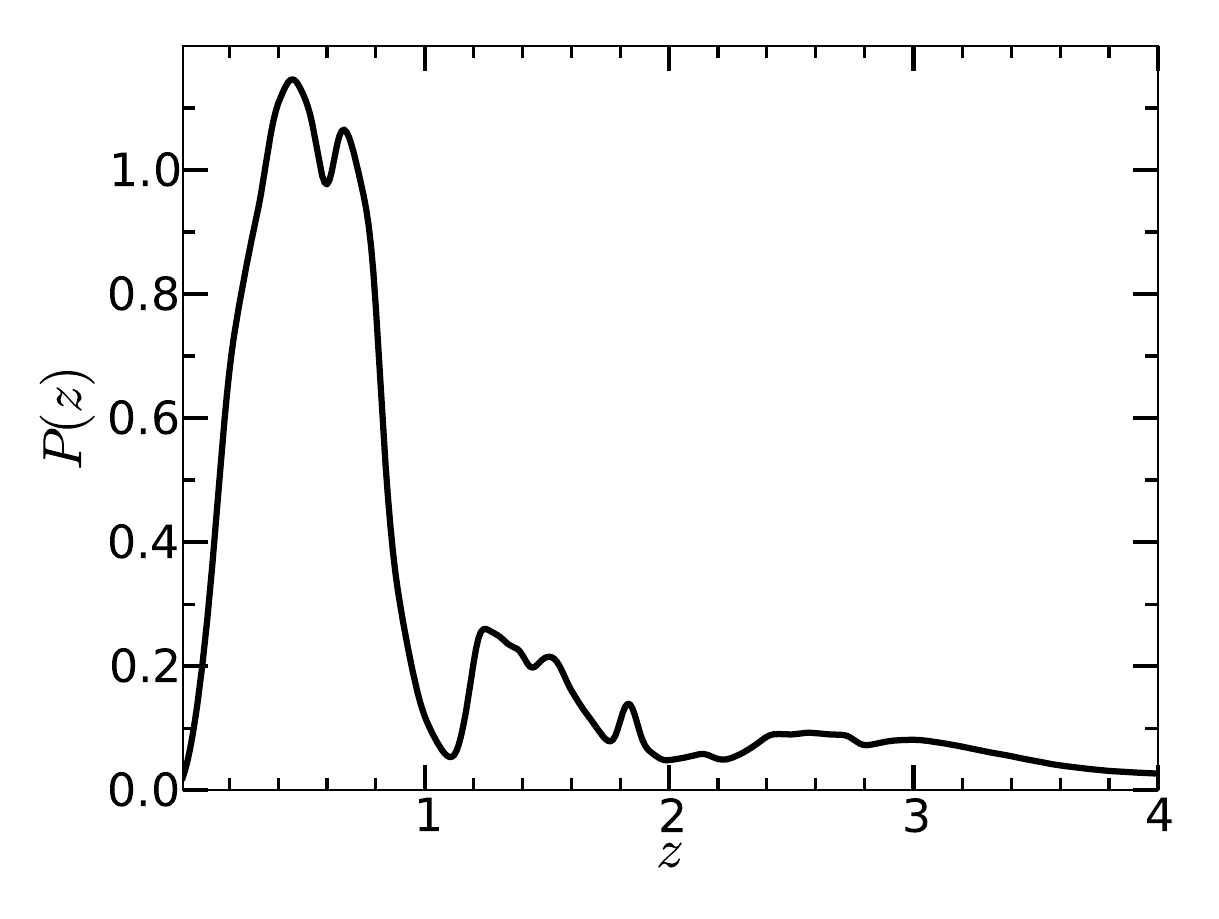}
\caption{The redshift probability distribution function for the third foreground galaxy (ID: 650), obtained with the photometric redshift code BPZ using the Hubble Ultra Deep Field prior. The distribution is reasonably broad and peaks around $z=0.2$--$0.4$, however a non-negligible peak at $z=1.2$--$1.4$ leads to the distinct possibility that this object may also lie at the same redshift as the nearby $z=1.327$ galaxies, forming a small group.}
\label{figure:p(z)_650}
\end{center}
\end{figure}

\indent The X-Shooter slit does not cover the third nearby foreground galaxy (ID: 650). To constrain the redshift of this source, we utilise the Bayesian photometric redshift code BPZ \citep{benitez2000bayesian, benitez2004faint, coe2007galaxies, benitez2008bayesian}. The photometric redshift distribution of this source is characterised by an extended probability distribution, peaking at $z\sim0.3$. However, there is a non-negligible secondary peak in the likelihood distribution centred at $z=1.3$, leading to the distinct possibility that this galaxy is at the same redshift as the central objects and the broad redshift distribution is simply an effect of photometric scatter. The photometric redshift distribution for this third galaxy is shown in Fig. \ref{figure:p(z)_650}.\\
\indent The possibility that the third foreground galaxy is part of the group with the two galaxies at $z=1.327$ supports a larger total dark matter mass of the deflector. However, the dark matter mass required for strong lensing in the lens modelling of the system (see Section \ref{section:MCMC}) is not ruled out if this additional galaxy is not part of the group. In Section \ref{section:MCMC} we model the system with only a single mass distribution centred on the centre-of-light of the spectroscopically confirmed galaxy pair, thereby assuming it resides at the centre of the halo of the group.\\
%

\section{Measurement of lens and source properties using a Markov-Chain Monte Carlo Bayesian inference}
\label{section:MCMC}
\indent In this section, we model the observed configuration using a singular isothermal ellipsoid (SIE) mass distribution for the foreground group of galaxies, centred at the centre of light of the spectroscopically confirmed galaxy pair, and a S\'{e}rsic brightness profile for the source. We have implemented the Metropolis algorithm in a Markov-Chain Monte Carlo analysis to find the posterior probability distribution function of source and deflector parameters within a Bayesian framework.\\
\indent Our deflector mass model consists of a SIE, which is a generalisation of the singular isothermal sphere (SIS), and is a commonly used parametrisation of the mass distribution of a lensing galaxy \citep{keeton2001catalog}. The SIE provides an adequate representation of the mass distribution of massive elliptical galaxies \citep{kochanek1995evidence, treu2004massive, koopmans2009structure, treu2010strong}. We adopt the definitions of the Einstein radius, ellipticity and position angle as described by \citet{keeton2001catalog}. The surface mass density, $\kappa$, in units of critical density, is defined by,
\begin{equation}
\kappa = \frac{1}{2}\frac{\theta_{ER}}{\sqrt{(1-\epsilon)x^{2} + (1+\epsilon)y^{2}}},
\end{equation}
where $\theta_{ER}$ is the Einstein radius, $\epsilon = \frac{1-q^{2}}{1+q^{2}}$ and $q$ is the axis ratio of the ellipsoid. According to this definition the Einstein radius is defined along the intermediate axis of the ellipsoid. The Einstein radius is then related to the velocity dispersion of the SIE profile by
\begin{equation}
\theta_{ER} = 4 \pi (\frac{\sigma_{\star}}{c})^{2} \frac{D_{\textrm{LS}}}{D_{\textrm{S}}},
\end{equation}
where $\sigma_{\star}$ is the line-of-sight stellar velocity dispersion, $D_{\textrm{S}}$ is the angular diameter distance from the observer to the source and $D_{\textrm{LS}}$ is the angular diameter distance from the deflector to the source. The centre of the SIE is fixed to coincide with the centre of light of the central galaxy pair. The deflector model therefore has three free parameters: Einstein radius (as a proxy for velocity dispersion/mass, and redshift), ellipticity and position angle.\\
\indent We adopt the S\'{e}rsic profile \citep{sersic1963influence} for the source surface brightness distribution. The source is defined by six free parameters: the S\'{e}rsic index, the effective radius, flux, $x$-position, $y$-position, ellipticity and position angle. This gives a total of 10 parameters for the deflector-source gravitational lens model.\\
\begin{figure}
\begin{center}
\includegraphics[trim=2 0 50 0, clip, scale=0.16]{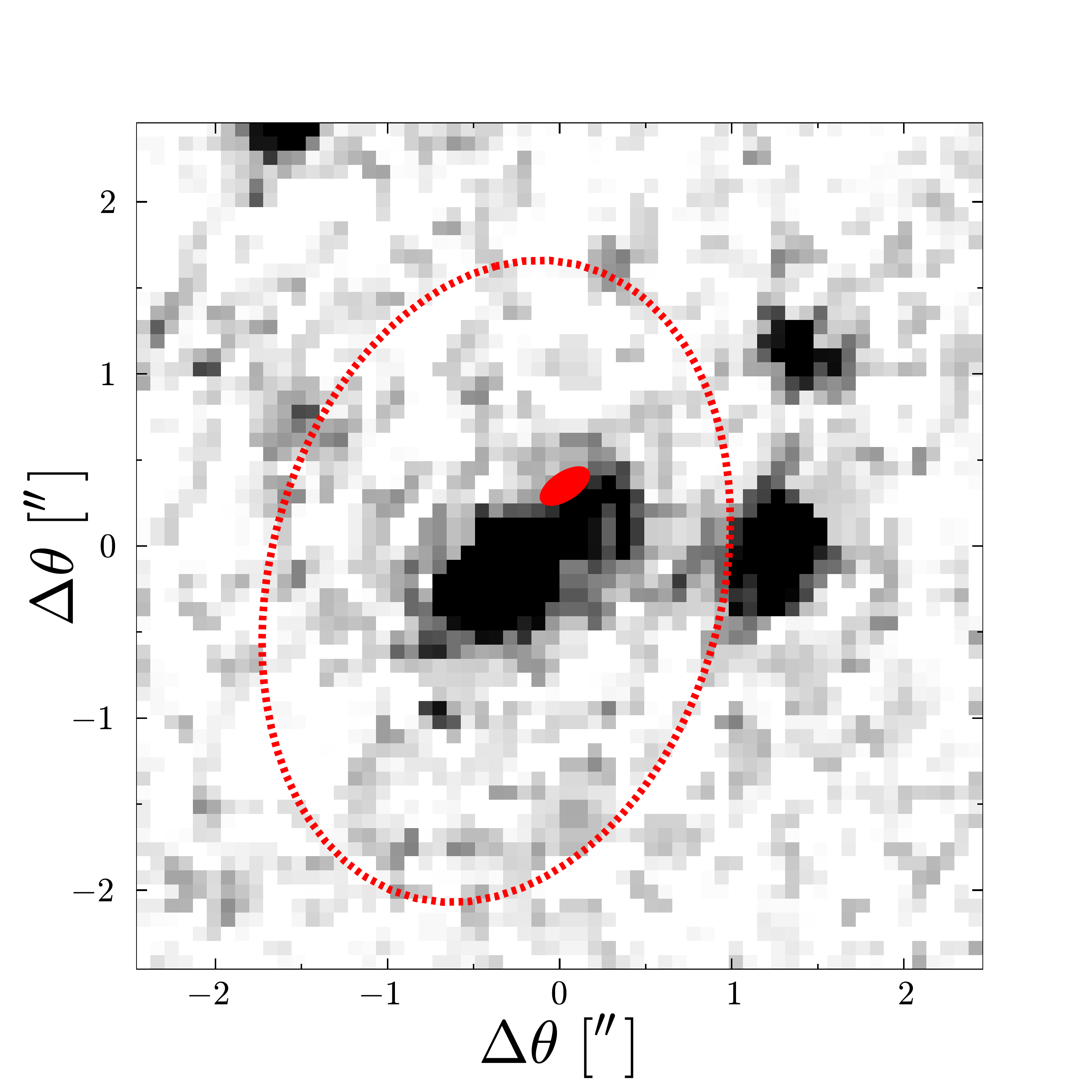}
\caption{The lens model with parameters at the median of their MCMC distribution. The red filled ellipse represents the source position in this model, and is shown with its intrinsic orientation and ellipticity. More detailed reconstructions of the source are shown in Fig. \ref{figure:source}. The critical curve is plotted (dashed red). This lens model accurately reproduces the positions of the primary and counter images. This model produces a flux ratio of $4.92$ consistent with the observed value. The tangential elongation in the primary image along the critical curve is evident in this figure.}
\label{figure:median-model}
\end{center}
\end{figure}

\subsection{The MCMC method and code}
\indent We wrote an MCMC sampler in $\tt{Python}$ using the Metropolis algorithm. The code compares the BoRG data with surface brightness models of the image plane produced using $\tt{GRAVLENS}$ \citep{keeton2001computational}.\\
\indent The code follows a standard MCMC process. The log-likelihood is $\frac{1}{2}\chi^{2}$, where $\chi^{2} = \sum\left(\frac{\textnormal{model}_{i} - \textnormal{data}_{i}}{\sigma_{i}}\right)^{2}$. The data are the WFC3 data in $H_{160}$ and $J_{125}$ and the model is the PSF-convolved image plane. We use uniform priors on all parameters. As lensing is achromatic, the only difference between the $H_{160}$ model and the $J_{125}$ model comes from the PSFs and the noise model. Hence, the log-likelihood is calculated for the $H_{160}$ and $J_{125}$ independently and summed to generate a total log-likelihood for a given model.\\
%
\begin{figure*}
\begin{center}
\includegraphics[trim=0 0 80 0, clip, scale=0.15]{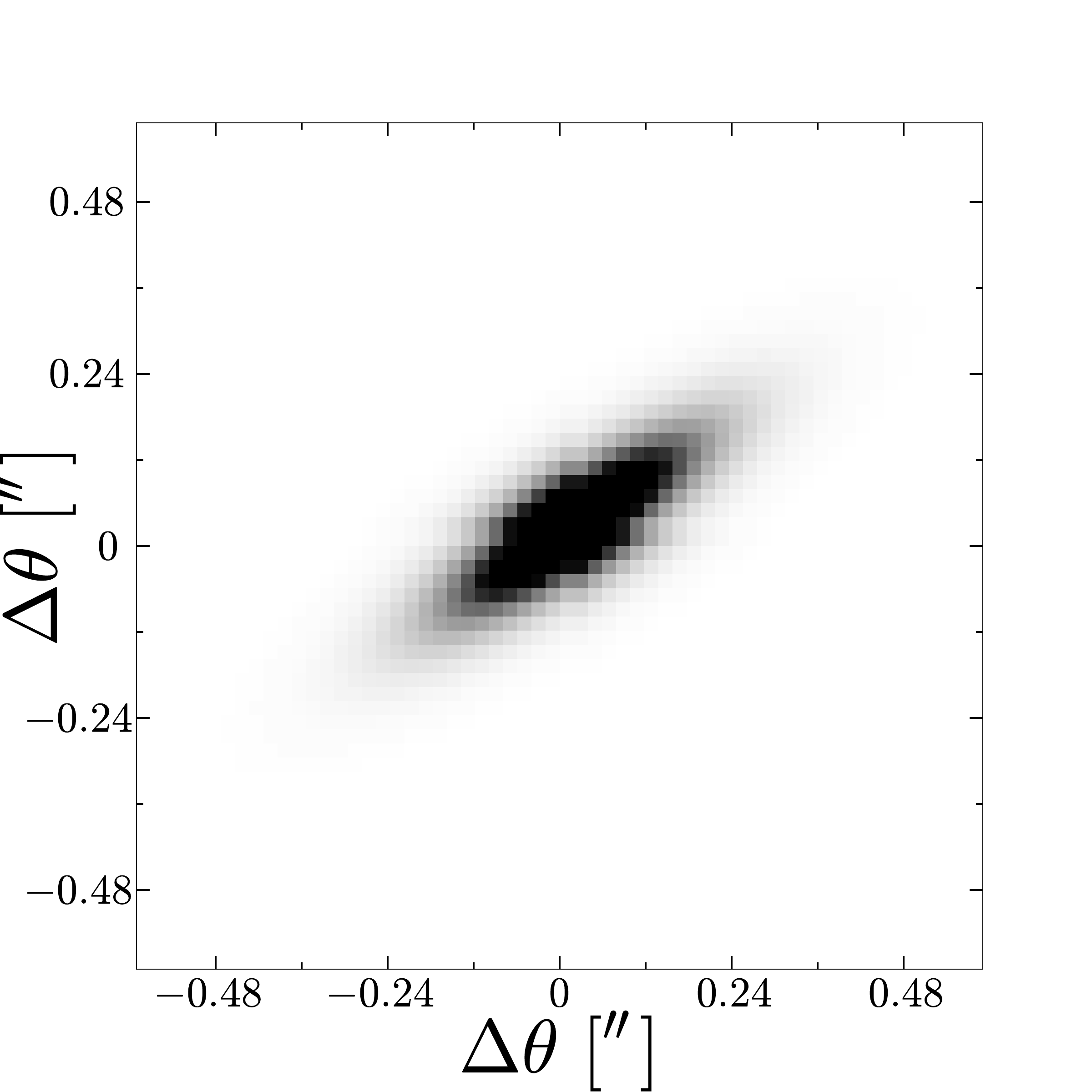}
\includegraphics[trim=0 0 80 0, clip, scale=0.15]{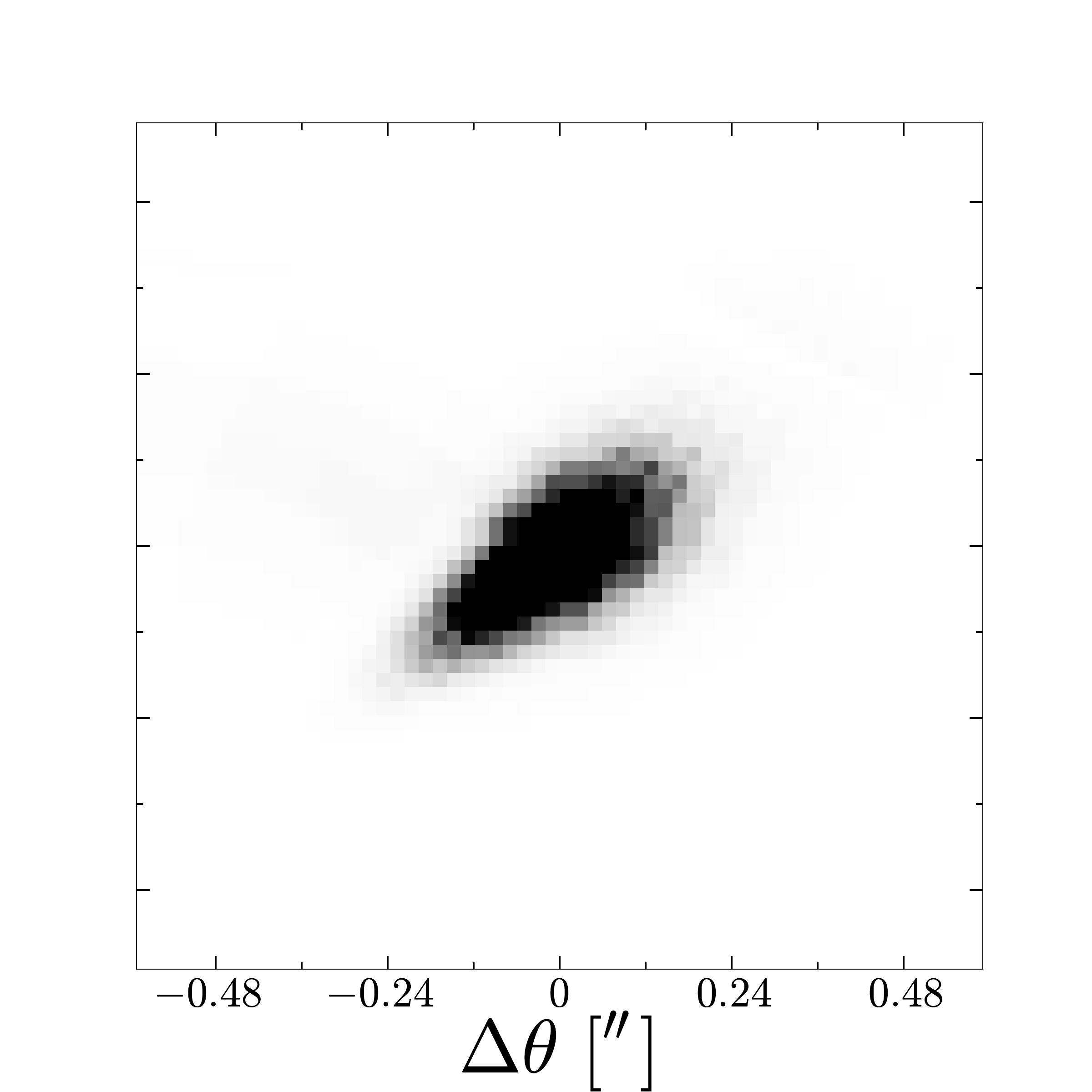}
\includegraphics[trim=0 0 30 0, clip, scale=0.15]{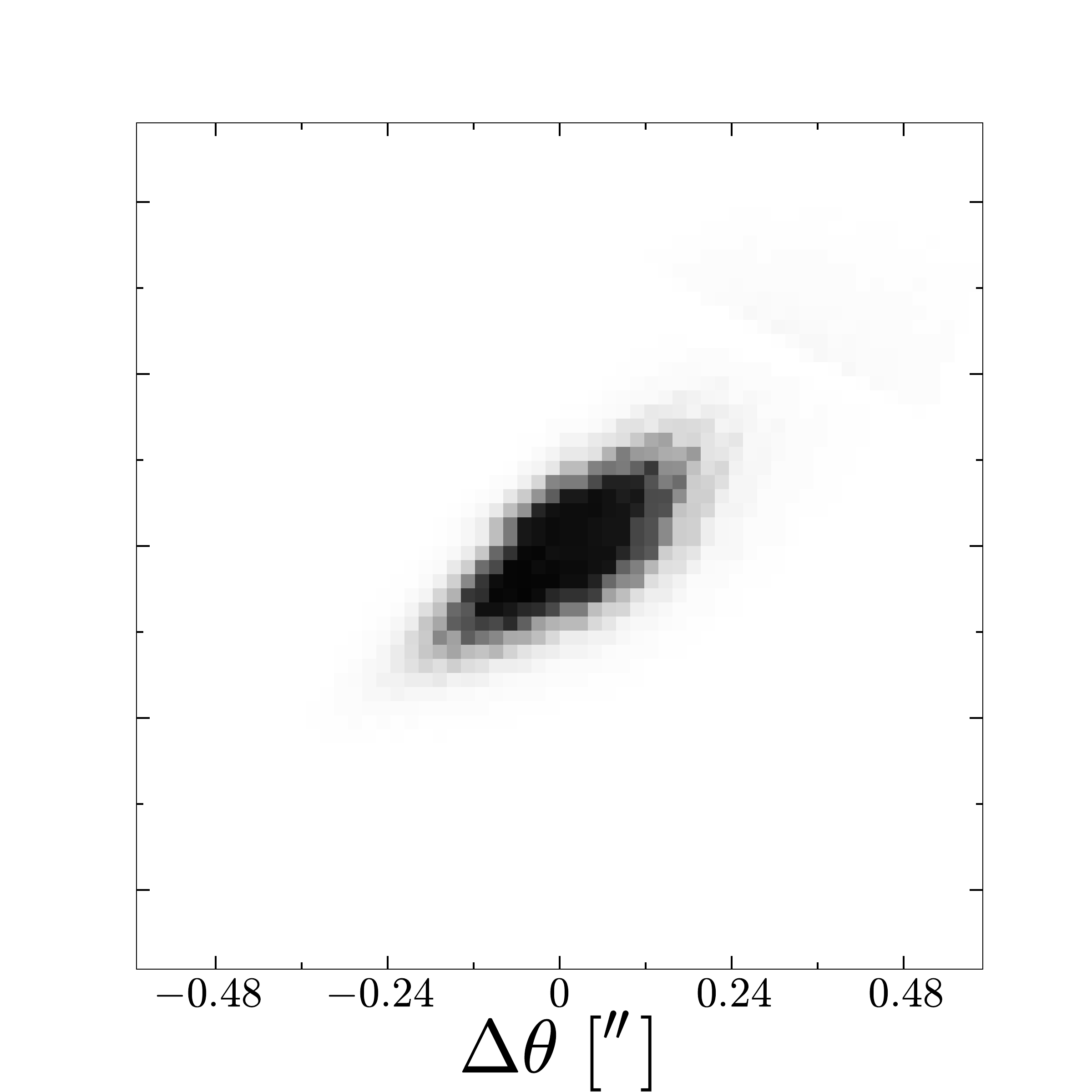}
\caption{Left: The MCMC recovered source. Centre: the $J_{125}$ $\tt{LENSVIEW}$ reconstruction. Right: the $H_{160}$ $\tt{LENSVIEW}$ reconstruction. As gravitational lensing magnifies surface area while preserving surface brightness, spatial information of the source ca be recovered with more detail than if it was not magnified. The source plane is sampled at a pixel scale $4$ times finer than the data. These sources have not been convolved with the HST WFC3 PSF. All three panels are $1.2$ arcsec on each side and shown on the same contrast scale.}
\label{figure:source}
\end{center}
\end{figure*}

\subsection{MCMC lens modelling}
\indent We run the MCMC code with the 10 parameters described above. The data are cropped WFC3 FITS files in $J_{125}$ and $H_{160}$, along with oversampled PSFs and root mean square (RMS) maps. The code allows for oversampled PSFs by oversampling the image plane equivalently. The step size is optimiSed to deliver a success rate of $1/3$. As is standard, we discard the MCMC burn in. We find well-constrained posterior probability distribution functions (posterior PDF) for each source and lens parameter.\\
\indent The inferred parameters from the MCMC analysis are presented in Table \ref{table:results}. A deflector with an Einstein radius of $1.49\pm0.03$ arcsec and ellipticity $\epsilon=0.70^{+0.05}_{-0.11}$ at $-15.3^{+2.7}_{-2.0}$ $^{\circ}$ East of North is favoured by the data. The source has a S\'{e}rsic index of $n=0.92^{+0.24}_{-0.16}$ and a major-axis half-light radius of $0.7\pm0.1$ kpc (in Section \ref{section:comparison} $\tt{SExtractor}$ is used to determine the circularised half-light radius for consistency with the literature). The apparent magnitude of the source would be $m_{J_{125}}=27.5\pm0.1$ mag if observed without magnification. The magnification at the location of the primary image is $\mu=4.3\pm0.2$. The total magnification is $\mu=5.2\pm0.3$. The lens model based on the median of the posterior PDF of each parameter is presented in Fig. \ref{figure:median-model}.\\
%
\begin{table}
\begin{center}
\begin{tabular}{| l  r |}
\hline
Parameter  & Posterior median\\
\hline
\hline
SIE $\theta_{\textrm{ER}}$			&$1.49\pm0.03$ arcsec\\
SIE ellipticity			&$0.31^{+0.03}_{-0.01}$\\
SIE position angle		&$-15.3^{+2.7}_{-2.0}$ $^{\circ}$\\
\hline	
Source x-position		&$0.39^{+0.02}_{-0.01}$ arcsec\\
Source y-position		&$0.54\pm0.01$ arcsec\\
Source flux				&$0.20^{+0.03}_{-0.02}$cps\\
Source ellipticity		&$0.70^{+0.05}_{-0.11}$\\
Source position angle	&$-57.0^{+2.8}_{-1.8}$ $^{\circ}$\\
Source $r_{\textnormal{e}}$		&$0.14\pm0.01$ arcsec\\
Source S\'{e}rsic index	&$0.92^{+0.24}_{-0.16}$\\
\hline
\end{tabular}
\caption{Posterior medians and 68 percent confidence intervals for the parameters of the deflector and source model. Uniform priors have been assumed for all parameters.}
\label{table:results}
\end{center}
\end{table}
\indent The source model is presented in the left panel of Fig. \ref{figure:source} at a finer pixel scale than the data as lensing allows the recovery of more spatial information than if the source was not lensed. The source is also reconstruct in the median lens model using $\tt{LENSVIEW}$ \citep{wayth2006lensview}. $\tt{LENSVIEW}$ creates a mapping matrix for a given lens model which reverse maps image plane pixels to the source plane. This is done independently for the $J_{125}$ and $H_{160}$ images. The $\tt{LENSVIEW}$ reconstructions are shown in the centre and right panels of Figures \ref{figure:source} and agree well with the MCMC results.\\
\indent This model predicts a flux ratio of the two images of $4.92$, which is consistent with the observed value of $4.8\pm2.6$. We did not impose an informative prior on the flux ratio, so all values are formally allowed in the model and deemed equally likely. Hence, a consistent predicted flux ratio with the measured flux ratio in Section \ref{subsection:counterphot} supports the reality of the counter image at its observed location. \\
\indent To investigate how constraining the marginal detection of the counter image is in the model, we run the MCMC code with the pixels at the location of the counter image masked. We find that in the absence of the counter image, the model is no longer well constrained and there is a degeneracy between the Einstein radius of the foreground galaxies, and the source position and magnification. This occurs because we place no informative priors on the Einstein radius and orientation of the foreground galaxies, hence massless galaxies and no magnification of the $z\sim8$ LBG is considered equally as likely as intermediate or strong magnification of the source. In order to constrain the magnification of the source in the absence of the counter image, we place a prior on the velocity dispersion of the foreground galaxies using the Faber-Jackson relation of \citet{barone2015impact}. Using this prior, we place a one sigma lower limit on the magnification of $\mu\geq1.38$ when masking the counter image pixels. \\
%

\section{Source morphology}
\label{section:comparison}
\indent In this section we compare the predicted source parameters with other $z\sim8$ galaxies from the literature. The intrinsic source profile recovered in our MCMC analysis, and independently using $\tt{LENSVIEW}$ on the $J_{125}$ and $H_{160}$ data, is presented in Fig. \ref{figure:source}. We use $\tt{SExtractor}$ on the intrinsic sources to determine their intrinsic effective radii. The effective radius definition is not unique in the literature. For consistency with previous work on galaxy sizes at high redshift \citep{grazian2012size} we use $\tt{SExtractor}$ to determine the effective radius, which uses concentric circular aperture measurements. In contrast, the effective radius value found in the MCMC analysis is defined along the major-axis of the S\'{e}rsic profile, which will vary by a factor of $\sqrt{q}$, where $q$ is the axis ratio. The intrinsic effective radii of the source, recovered in the MCMC analysis and $\tt{LENSVIEW}$, are presented in Table \ref{table:sizes}, and are consistent with the results of \citet{oesch2009structure}, who found the average intrinsic size of a $z\sim8$ LBG to be $\sim 0.5$kpc.\\
\indent PSF broadening affects the measurement of effective radii. To first approximation, 
\begin{equation}
r^{\textnormal{obs}}_{e}= \sqrt{r^{\textnormal{intr}^{2}}_{\textnormal{e}} + r^{2}_{\textnormal{PSF}}}
\end{equation}
where $r^{\textnormal{obs}}_{\textnormal{e}}$ is the observed effective radius, $r^{\textnormal{intr}}_{\textnormal{e}}$ is the intrinsic effective radius and $r_{\textnormal{PSF}}$ is the full width at half maximum of the PSF. Table \ref{table:sizes} summarizes the measurements of the observed size.\\

\begin{figure}
\begin{center}
\includegraphics[scale=0.4]{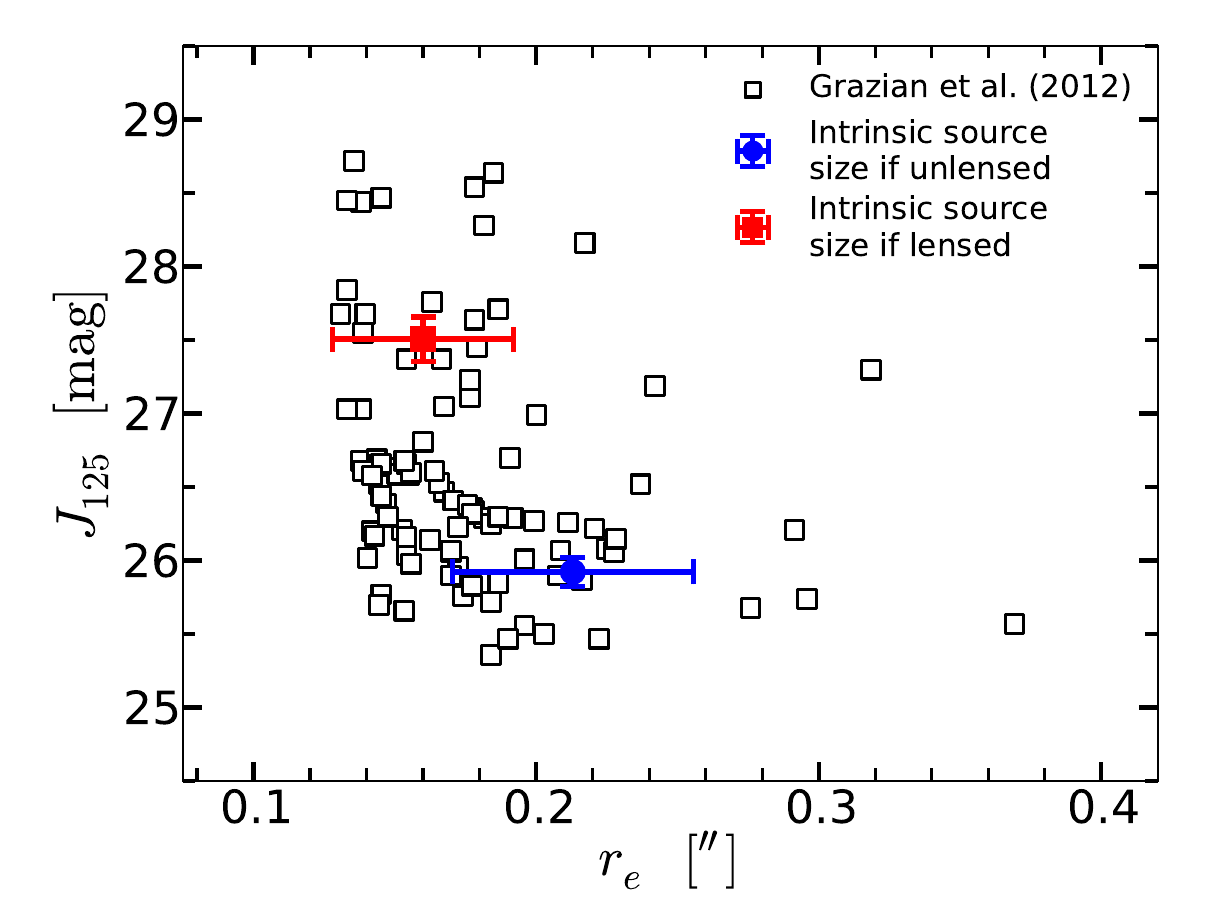}
\caption{Size versus luminosity relation for the $Y_{098}$-dropout source under the lensing scenario, as recovered from our MCMC analysis and as reconstructed with $\tt{LENSVIEW}$ (red, $m_{J}=27.5$ mag). The blue point ($m_{J}=25.9$) shows the same measurement but assuming that no lensing magnification is present. Both scenarios are consistent with the size-luminosity relation measured by \citet{grazian2012size} at $z\sim 7$, and scaled to $z\sim 8$ assuming $r(z) \propto 1/(1+z)$ \citep{oesch2009structure, wyithe2011extrapolating}. The change in distance modulus from $z=7$ to $z=8$ is accounted for.}
\label{figure:size-luminosity}
\end{center}
\end{figure}

\indent We compare our results with the size-luminosity relation measured at $z\sim7$ by \citet{grazian2012size} in Fig. \ref{figure:size-luminosity}. We account for an expected approximate scaling of the galaxy radius with redshift of $1/(1+z)$, in line with the study by \citet{oesch2009structure} who constrained the redshift evolution of the scale-length of galactic disks in the HUDF, and theoretical predictions by \citet{wyithe2011extrapolating}. We find that the recovered properties of our source are consistent with the $z\sim7$ sample presented in that work. The $\tt{LENSVIEW}$ reconstruction and the source recovered in the MCMC analysis are plotted in red in Fig. \ref{figure:size-luminosity}. We also plot the size and luminosity of the primary image if it has not been lensed and find that the galaxy would not be an outlier on the relation. \\
\indent Additionally, we compare our results with the size-luminosity relations of \citet{shibuya2015morphologies} and \citet{curtis2014no}. \citet{shibuya2015morphologies} find the size of $\sim L_{\star}$ LBGs at $z\sim8$ to be $0.419^{+1.981}_{-0.262}$kpc, consistent with the LBG if it has not been significantly magnified, and the size of LBGs two magnitudes fainter to be $0.243^{+0.225}_{-0.068}$kpc, consistent with the size of the LBG under the strong lensing hypothesis. We also find that the size of the LBG under the lensing hypothesis is consistent with the sizes of the $z\sim8$ LBGs in \citet{curtis2014no}, but also consistent with the absence of lensing due to the scatter in the relation. Thus, the size of the $Y_{098}$-dropout is consistent with literature data whether the strong lensing hypothesis is true or not. \\
\begin{table*}
\begin{center}
\begin{tabular}{| l  c  c  c  c |}
\hline
Source & Angular $r^{\textnormal{intr}}_{\textnormal{e}}$ [arcsec] & Proper $r^{\textnormal{intr}}_{\textnormal{e}}$ [kpc] & Angular $r^{\textnormal{obs}}_{\textnormal{e}}$ [arcsec]& Proper $r^{\textnormal{obs}}_{e}$ [kpc]\\
\hline
\hline
MCMC analysis 				& $0.09\pm0.02$	& $0.45\pm0.08$	& $0.16\pm0.03$	& $0.80\pm0.15$\\
$J_{125}$ $\tt{LENSVIEW}$	& $0.10\pm0.02$	& $0.50\pm0.09$	& $0.16\pm0.03$	& $0.80\pm0.15$\\
$H_{160}$ $\tt{LENSVIEW}$	& $0.10\pm0.02$	& $0.50\pm0.09$	& $0.17\pm0.03$	& $0.85\pm0.16$\\
\hline
\end{tabular}
\caption{The intrinsic and WFC3 observed effective radii of the source recovered in the multi-band MCMC analysis, and reconstructed independently for each band using $\tt{LENSVIEW}$. The intrinsic values are the source sizes once corrected for PSF-broadening, while the observed sizes include the effect of PSF-broadening.}
\label{table:sizes}
\end{center}
\end{table*}
%

\section{Deflector mass-to-light ratio}
\label{section:M/L}
\indent Strong gravitational lensing offers a direct and uniquely precise measurement of the deflector mass distribution. Lensing constraints are free from assumptions of a system's dynamical properties which is critical in other direct measurements of a system's mass \citep{cappellari2009dynamical}. Under the assumption that we are, in fact, observing a lensing system as described in this paper, we can measure the mass of a $z=1.327$ group and its associated massive galaxies with unprecedented precision. We measure the mass enclosed within the Einstein radius of the deflector system, as well as the total mass enclosed within the effective radius of the brighter galaxy of the spectroscopically confirmed deflector.\\
\indent To calculate the mass enclosed by the Einstein radius, we integrate the SIE mass distribution over the Einstein radius of $1.49$ arcsec. We find the mass within the effective radius by fitting a de Vaucouleurs profile to the light from the main deflector galaxy using $\tt{GALFIT}$ in order to determine its effective radius, $r_{\textnormal{e}}$. The mass within the effective radius is then calculated by integrating the SIE mass distribution over $r_{\textnormal{e}}$. It should be noted that the effective radius of the de Vaucouleurs profile is defined by $\tt{GALFIT}$ along the major axis. For consistency with the mass measurement within the Einstein radius, we integrate the mass over the effective radius defined along the intermediate axis.\\
\indent Calculating the physical masses and intrinsic luminosities of the deflector galaxies requires knowledge of the source and deflector redshifts. We consider the uncertainty in the spectroscopic redshift of the main deflector to be negligible compared to the uncertainty in the source LBG's redshift, and other systematic errors. The non-negligible range for the redshift PDF of the source redshift is $7.6\lesssim z \lesssim8.8$, and the uncertainty on the brighter galaxy's effective radius is approximated to be $20$ per cent in order to account for systematic errors which dominate the formal statistical errors provided by $\tt{GALFIT}$. The centre of the deflector is fixed in the MCMC analysis, but we account for uncertainty in its location when calculating the physical mass enclosed within the effective radius of the brighter galaxy. We also include errors on the solar luminosity, K correction, systematics in the photometry and random errors in the MCMC analysis. The resulting values for measured quantities are quoted at their median and the resulting errors on the derived quantities are quoted and plotted at the $68$ per cent confidence limit.\\
\indent The mass enclosed within the Einstein Radius of $1.49$ arcsec of the SIE mass distribution is $(9.62\pm0.31)\times10^{11} M_{\odot}$. Photometry of the luminous matter enclosed within the Einstein radius is listed in Table \ref{table:photometry}. We apply the distance modulus out to $z=1.327$ and a K-correction using single stellar populations models with ages in the range 1-2 Gyr to convert the $J_{125}$ magnitude to the rest-frame $B$-band magnitude. We note that the redshifted B band is very close to the observed $J_{125}$ and therefore the K correction is fairly robust with respect to changes in the assumed spectral energy distribution, resulting in a negligible additional uncertainty. Combining these measurements, we find a total mass-to-light ratio within the Einstein radius of $M/L_{B}=38.8^{+4.2}_{-3.6} M_{\odot}/L_{\odot}$. This suggests that dark matter dominates the mass of the deflector and supports the argument that the deflector is a group. The Einstein radius is $\sim20$ times as large as the effective radius of the brighter deflector galaxy, which is consistent with a mass distribution dominated by dark matter.\\
%
%
\indent To obtain the stellar mass-to-light ratio of the brighter central deflector, we calculate the lensing mass within its effective radius and account for the dark matter fraction as follows. We use $\tt{GALFIT}$ to fit a de Vaucouleurs profile to the main deflector galaxy. We find it to have circularised $r_{\textnormal{e}}=0.072\pm0.014$ arcsec, and a magnitude of $m_{\textnormal{deVauc}} = 23.55\pm0.1$ mag. We find the circular effective radius to be $0.62^{+0.14}_{-0.13}$kpc, making the galaxy very compact and less luminous than the most massive galaxies at comparable redshift \citep{damjanov2009red}. Within this galaxy's effective radius the SIE mass distribution has a mass of $4.3\pm0.3\times10^{10} M_{\odot}$. The left hand panel of Fig. \ref{figure:m2L} shows this galaxy's stellar mass and effective radius in comparison with the sample of elliptical galaxies at $1<z<1.5$ from \citet{newman2012can}.\\
\indent We apply the distance modulus and K correction to convert the $J_{125}$ magnitude to the rest-frame $B$-band magnitude. We find that the mass-to-light ratio within its effective radius is $M/L_{B} = 3.5^{+1.2}_{-0.9} M_{\odot}/L_{\odot}$. Adopting the average dark matter fraction of $1/3$ found for SLACS systems of this mass \citep{treu2004massive,auger2010dark}, we obtain a stellar mass-to-light ratio of $M_{\star}/L_B=2.3^{+0.8}_{-0.6} M_{\odot}/L_{\odot}$. We find that the brighter deflecting galaxy is consistent in size and mass with compact, massive galaxies observed at similar redshifts \citep[so-called `red nuggets',][]{daddi2005passively, van2008confirmation, damjanov2009red, newman2012can}. The right hand panel of Fig. \ref{figure:m2L} compares our measurement of the stellar mass-to-light ratio to previous measurements of the cosmic evolution of the stellar mass-to-light ratio obtained using lensing measurements at lower redshifts \citep{treu2004massive} and via the evolution of the Fundamental Plane out to $z\sim1.2$ \citep{treu2002evolution}.\\

\section{Discussion}
\label{section:discussion}
\indent In this paper, we have analysed what appears to be a case of strong gravitational lensing of a $z\sim8$ LBG found in the BoRG survey. In the event that this source has been strongly lensed, we find the empirical strongly lensed fraction of LBGs detected at $8\sigma$ in BoRG to be $10$ per cent. We discuss the supporting evidence for this hypothesis in this section.\\

\subsection{Previous studies of the lensed fraction of high-\lowercase{$z$} LBGs}
\indent The study of $z\sim7$--$8$ LBGs in the XDF and CANDELS blank fields found the strongly lensed fraction of LBGs brighter than $M_{\star}$ to be $\sim6$ per cent \citep{barone2015impact}. In a separate study by \citep{mason2015correcting}, the strongly lensed fraction in the BoRG fields is estimated at $3$--$15$ per cent. These studies were based on the work by \citet{wyithe2011distortion}, who predicted a strongly lensed fraction of 10 per cent for high redshift samples with a flux limit of $M<M_{\star}$. These studies lead to the expectation that $\approx1$ of the 10 high-S/N LBGs, which are all approximately $M_{\star}$ or brighter, in the BoRG sample will show evidence for strong gravitational lensing.
\begin{figure*}
\begin{center}
\includegraphics[scale=0.42]{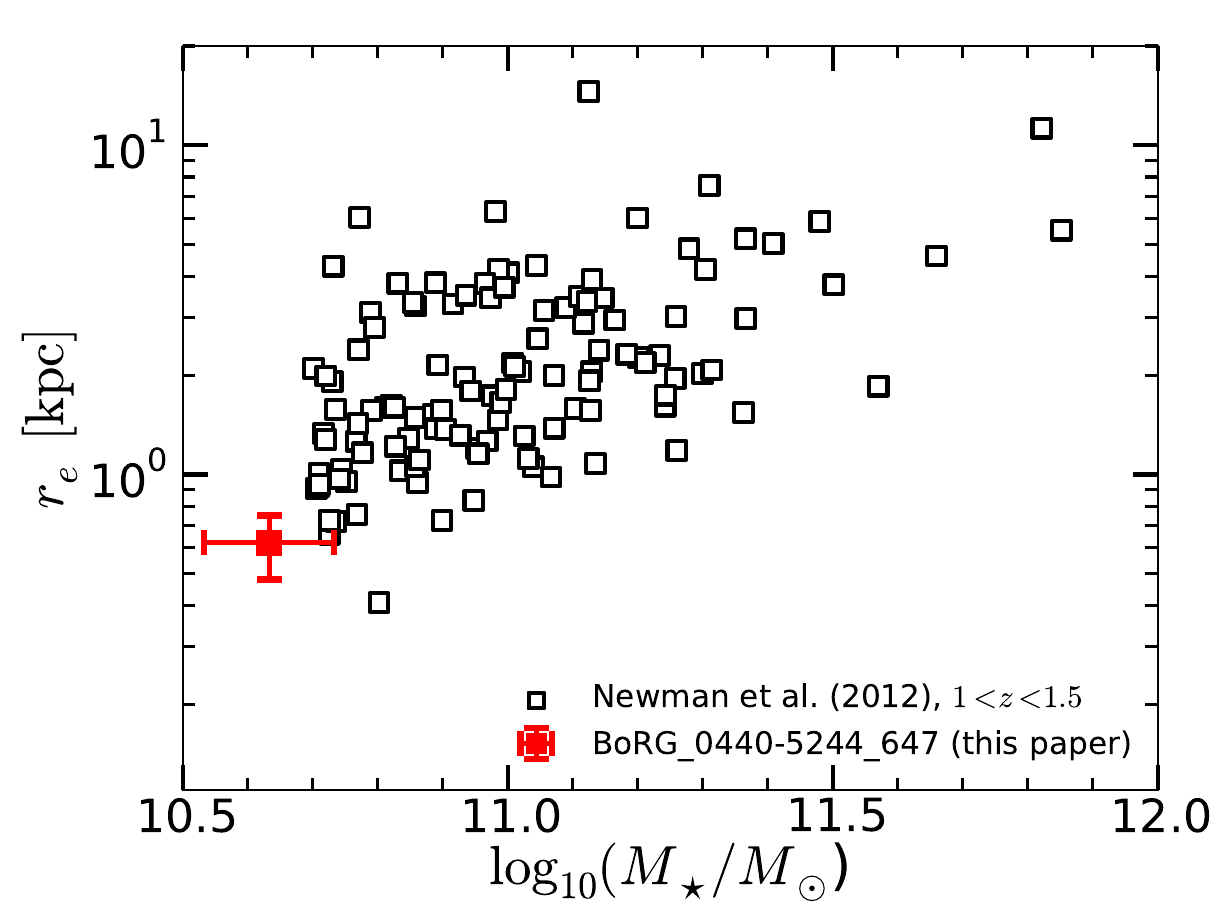}
\includegraphics[scale=0.42]{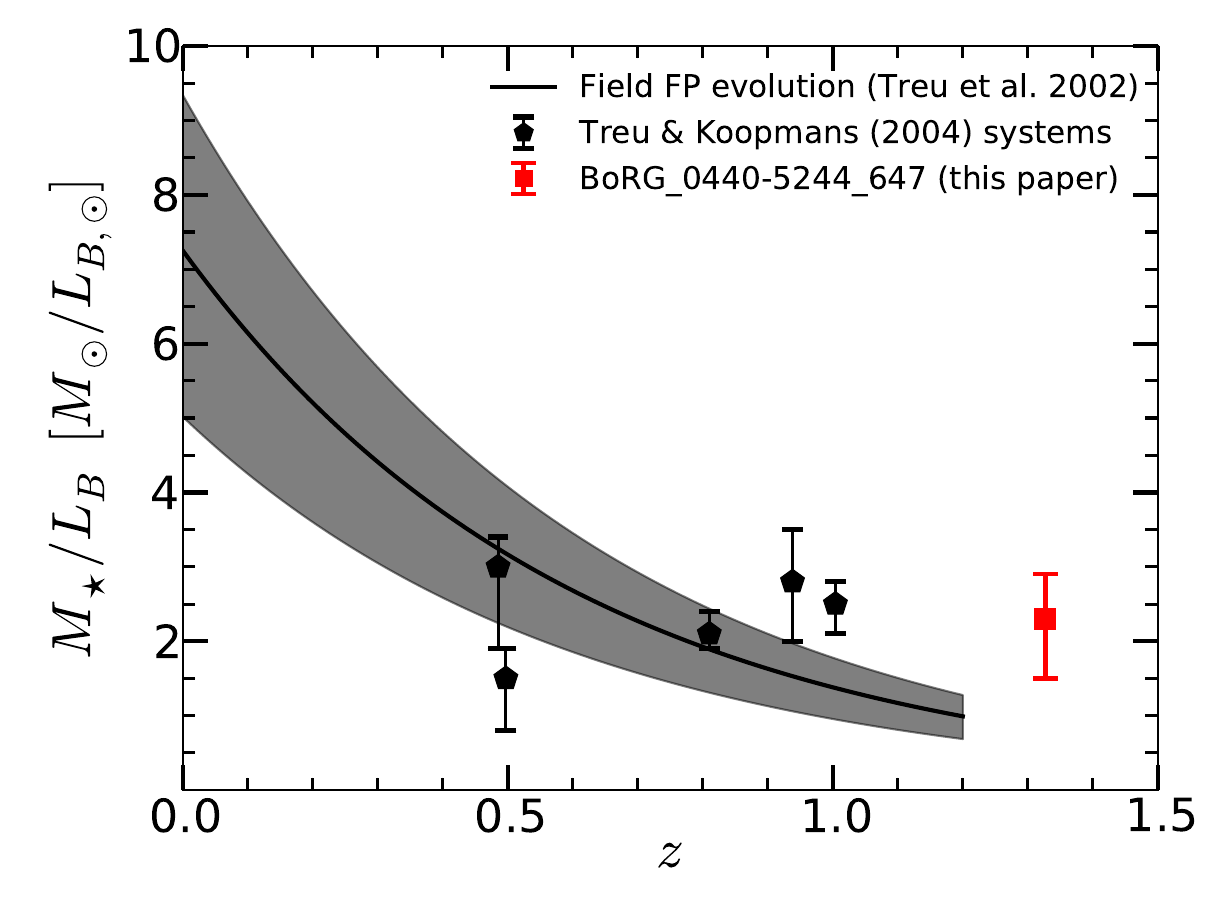}
\caption{Left: The size mass relation at $1<z<1.5$. The open squares show the measured stellar masses from spectral synthesis models of multicolor photometry \citep{newman2012can}. Our lens candidate (red point) falls on the relation, using our lensing-based estimate of the stellar mass presented in this work. Right: Cosmic evolution of the stellar mass-to-light ratio based on gravitational lenses at lower redshift \citep{treu2004massive}, the evolution of the Fundamental Plane measured out to $z\sim1.2$ \citep{treu2002evolution}. The red point represents the candidate lens presented in this paper.}
\label{figure:m2L}
\end{center}
\end{figure*}

\subsection{Proximity to bright foreground objects}
\begin{figure}
\begin{center}
\includegraphics[trim=15 0 0 15, clip, scale=0.42]{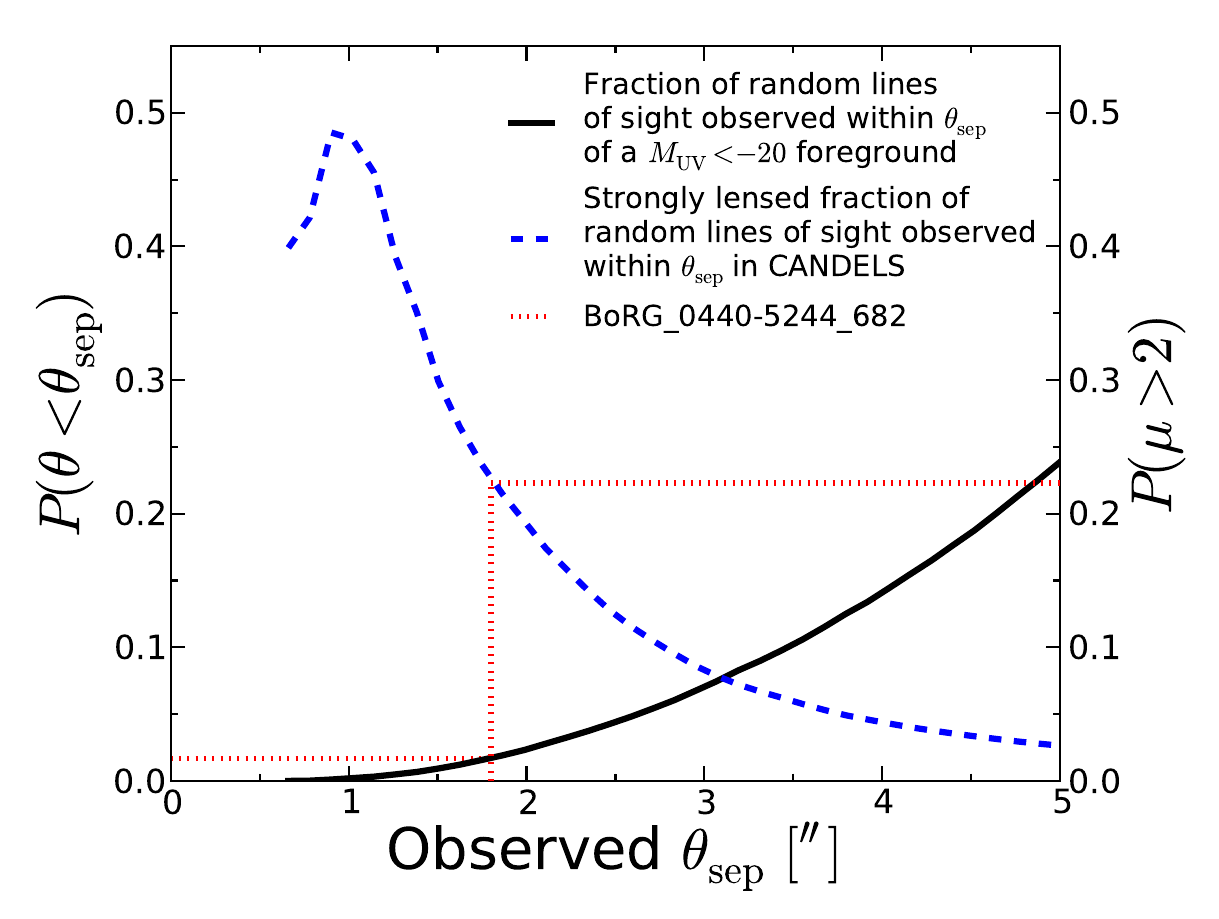}
\caption{The probability of observing a random $z\sim8$ source within $\theta_{\textrm{sep}}$ of a foreground galaxy with $M_B<-20$ mag (solid black), and the likelihood of strong lensing, $P(\mu>2)$, of random lines of sight observed within $\theta_{\textrm{sep}}$ of a foreground galaxy with $M_B<-20$ mag (dashed blue) in the CANDELS fields using the model of \citet{barone2015impact}. The red dotted lines mark the observed separation between borg$\_0440$-$5244\_682$ and borg$\_0440$-$5244\_647$ ($1.8$ arcsec). The likelihood of this random alignment is $\sim1.7$ per cent, and the likelihood of strong lensing is $\sim23$ per cent.}
\label{figure:proximity}
\end{center}
\end{figure}

\indent The $z\sim8$ LBG we have identified as a good candidate for being strongly gravitationally lensed, borg$\_0440$-$5244\_682$, is observed in close proximity to multiple bright foreground objects, so there is a considerably enhanced chance of strong lensing for this LBG. Analysis of the CANDELS fields from \citet{barone2015impact} shows that random source positions observed within $\sim1.8$ arcsec of a foreground galaxy with $M_{B}<-20$ mag have a greatly enhanced likelihood of being strong lensed over all random source positions. This analysis shows that the fraction of 50 000 random source positions found to be projected nearby to bright foreground objects is low ($2$ per cent of all random source positions are observed within $2.0$ arcsec of a bright foreground galaxy). But of the $2$ per cent which are projected within $2.0$ arcsec, $\sim20$ per cent will have been strongly gravitationally lensed.\\ 
\indent Fig. \ref{figure:proximity} shows the likelihood of a random line of sight being observed within $\theta_{\textrm{sep}}$ of a foreground galaxy with $M_{B}<-20$ mag (solid black), and the likelihood of strong lensing as a function of $\theta_{\mathrm{sep}}$\footnote{This analysis is based on observations of the CANDELS fields using the methodology of \citet{barone2015impact}.}. An object being observed within $1.8$ arcsec of a bright foreground objects implies that it has a $23$ per cent chance of being strongly lensed. The probability of strong lensing is even more enhanced for objects observed around brighter objects (e.g. $M_B<-20.5$ mag), giving a likelihood of lensing of $\sim30$ per cent at a separation of $\theta_{\mathrm{sep}}=1.8$ arcsec. The central galaxy pair (identified as a single foreground object), borg$\_0440$-$5244\_647$, has an absolute magnitude of $M_B=-20.5$ mag.\\
\indent Bayes Theorem gives a similar result, where the likelihood of strong lensing (``SL'') given the observed separation between a high redshift object and a foreground object can be expressed as,
\begin{equation}
P(\textrm{SL}|\theta\leq1\farcs8)=\frac{P(\theta\leq1\farcs8|\textrm{SL})}{P(\theta\leq1\farcs8)}P(\textrm{SL})
\end{equation}
where we can estimate values for $P(\textrm{SL})$ and $P(\theta\leq1\farcs8|SL)$ from fig. 4 of \citet[][]{barone2015impact} and $P(\theta\leq1\farcs8)$ from Fig. \ref{figure:proximity}, and find that,
\begin{align}
P(\textrm{SL}|\theta\leq1\farcs8)&\simeq\frac{0.6}{0.017}\times0.008\\
&=28\%.
\end{align}

\subsection{The foreground objects are typical deflectors of $\lowercase{z}\sim8$ LBGs}
\indent In the study of gravitational lensing in the CANDELS fields by \citet{barone2015impact}, they presented the empirical distribution of deflector luminosities, redshifts and deflector-source separations \citep[see fig. 4 of][]{barone2015impact}. The spectroscopic redshift of the central foreground galaxy pair in this paper is consistent with the observed peak of the redshift distribution for gravitational lenses of $z\sim7$-$8$ LBGs, which occurs between $z\sim1$-$2$. The observed separation between the central foreground objects and the $z\sim8$ LBG candidate, $\theta_{\textrm{sep}}=1.8$ arcsec, is also at the peak of the deflector-source separation distribution, which peaks at $1.5<\theta_{\textrm{sep}}<2.0$ arcsec. The luminosity of the central object is also near the peak of the absolute magnitude distribution. Additionally, the required Einstein radius derived in Section \ref{section:MCMC} of $\sim1.5$ arcsec is consistent with strong lensing by a massive galaxy/group at moderate redshift \citep{treu2002internal,treu2010strong,newman2015luminous}.\\
\indent Further to the point of the objects borg$\_0440$-$5244\_647$ and borg$\_0440$-$5244\_650$ being in the range of luminosities of gravitational deflectors in the CANDELS fields identified by \citet{barone2015impact}, we can also infer the mass of these galaxies using abundance matching. Based on modelling of the galaxy luminosity function at high redshift \citep{trenti2014luminosity}, abundance matching of the luminosity of the object borg$\_0440$-$5244\_647$ translates to an halo mass of $\sim 5.7\times 10^{12}~\mathrm{M_{\sun}}$. The object borg$\_0440$-$5244\_650$'s luminosity translates to a halo mass of $\sim 1.0\times 10^{12}~\mathrm{M_{\sun}}$ if it is also at $z\sim1.3$.\\
\indent Associating these masses with the virial mass, and circular velocities with the virial velocity, this mass would would give circular velocities $V_c\sim 270$ kms$^{-1}$ and $V_c\sim 150$ kms$^{-1}$ for the halos of the two galaxies borg$\_0440$-$5244\_647$ and borg$\_0440$-$5244\_650$, respectively. Translating circular velocities into central stellar velocity dispersions $\sigma_{\star}$ of an isothermal model for the deflector is complicated by baryonic physics which generally tend to increase the central velocity dispersion of a collapsed \citet{navarro1997universal} dark matter halo. In practice, the circular velocity is comparable to the observed stellar velocity dispersion $\sigma_{\star} \approx V_c$ at the scale considered here \citep[e.g.,][]{bundy2007mass,dutton2010kinematic}, which in turn is very close to the velocity dispersion of the singular isothermal sphere that describes the total mass density profile in the inner regions of massive elliptical galaxies \citep{treu2004massive,koopmans2009structure,auger2010sloan}. In addition, the gravitational lensing optical depth is a strongly rising function of velocity dispersion (e.g. \citealt{ruff2011sl2s}) and $\sigma_{\star}\approx 300$kms$^{-1}$ is a typical value for group-sized halos \citep{treu2002internal}. These inferred velocity dispersions are sufficient to produce an Einstein radius of the magnitude required in Section \ref{section:MCMC}.

\subsection{Possible multiple imaging}
\indent Multiple imaging by a deflector is a necessary condition for strong lensing, although secondary images of the source will appear much fainter and closer to the deflector galaxy, making them difficult to detect at high significance \citep{wyithe2011distortion, fialkov2015distortion}. Consistent with this, we observe a possible faint counter image on the opposite side of the central deflector objects. Both the brighter image and the fainter image are completely absent in the $Y_{098}$ and $V_{600}$ filters, consistent with the $z\sim8$ LBG selection criteria, and with the lensing hypothesis, as gravitational lensing is achromatic. The colours of the counter image are consistent with both the $Y_{098}$-dropout, and being a $z\sim8$ object below the flux limit of the BoRG field. The observed flux ratio between the two images is consistent with the flux ratio predicted in the best fitting model in our lens modelling. \\
\indent As discussed in Section \ref{subsection:counterphot}, the counter image is not detected in the $H_{160}$ image. This is expected based on the colours of the source and the sensitivity of the $H_{160}$ and $J_{125}$ images. \\

\subsection{Elongation of the dropout}
\indent The $z\sim8$ LBG is consistent with being elongated tangentially to the direction to the central foreground galaxies. In gravitational lensing theory, images are elongated into arcs and arclets along the critical curves of deflectors. Unless the deflector has very high ellipticity, the image's elongation will always be tangential to the direction from the image to the deflector. As described in Section \ref{subsection:photometry}, the primary image shows elongation that is tangential to the direction towards the central foreground galaxies with an ellipticity of $\epsilon=0.4\pm0.1$. Our modelling of the system confirms that the observed elongation is consistent with that predicted by the model once smearing by the WFC3 PSF and low S/N of the data are taken into account. \\

\subsection{Lack of companion dropouts in the field}
\indent Finally, we note that the field contains only this single candidate detected at high S/N. Based on dark matter (DM) clustering, we expect additional dropouts in the field if the primary image is not strongly magnified. Abundance matching predicts that the dropout is hosted in a rare dark matter halo with $M_{\textnormal{DM}}\gtrsim 2\times 10^{11}M_{\odot}$ \citep{munoz2008verifying,trenti2012overdensities}. However, the connection between DM halo mass and galaxy luminosity seems unclear at high redshift, as highlighted by the recent measurement of the two-point correlation function at $z\gtrsim 7$ from CANDELS and XDF catalogs \citep{barone2014measurement} and by the lack of strong clustering around a more luminous and isolated galaxy at $z=7.7$ \citep{oesch2015spectroscopic}. Therefore, while the lensing hypothesis provides a natural explanation of this single bright candidate (the dropout would be intrinsically fainter and hence less clustered), further studies are needed to use the lack of clustering as a quantitative support for the lensing hypothesis.\\

\subsection{Comparison with previous lensing studies of the BoRG sample}
\label{subsection:vdisps}
\indent The configuration we have investigated in this paper was not identified in the analysis of \citet{mason2015correcting} as a strong lensing candidate. Based on a Faber-Jackson relation \citep{faber1976velocity}, \citet{mason2015correcting} only identified a single different object from our high S/N sample as a candidate for being gravitationally magnified by a foreground object (inferring a magnification for that object of $\mu=1.47\pm0.3$). However, there are a number of reasons that the galaxy we have identified in this paper was missed by \citet{mason2015correcting}. Firstly, we have obtained additional spectroscopy which places the nearby objects at a lower redshift than the photometric redshift used in the \citet{mason2015correcting} analysis, which leads to a lower required mass for strong gravitational lensing. Secondly, the Faber-Jackson relation used by \citet{mason2015correcting} will underestimate the lensing potential of a gravitational deflector which consists of multiple objects, because gravitational magnification is not linearly additive \citep{wong2012optimal} and much more dark matter is expected in group environments.\\
\indent We find that when the central foreground galaxy pair, which is identified as a single object in the BoRG catalogue, is considered at its correct redshift, the estimated magnification of borg$\_0440$-$5244\_682$ in the \citet{mason2015correcting} framework increases by $\sim10$ per cent to $\mu=1.41$. We also infer the magnification with velocity dispersions found using the Faber-Jackson relation of \citet{barone2015impact} and their lensing likelihood framework, which leads to the estimated magnification of borg$\_0440$-$5244\_682$ to be $\mu=1.47^{+0.30}_{-0.18}$, and a strong lensing likelihood of 11 per cent. \\
\indent If the object borg$\_0440$-$5244\_650$ also resides at $z=1.3$ and is included as a third deflector, the magnification further increases at the location of the dropout. In fact, when including this third lensing galaxy at $z\simeq 1.3$ and using the velocity dispersions inferred from the FJR of \citet{barone2015impact}, the magnification at the location of the dropout is $\mu=1.96^{+0.63}_{-0.36}$, giving it a strong lensing likelihood of 47 per cent.\\

\indent The case presented in this paper highlights that the available photometry of both the source LBG and the photometric and spectroscopic observations of the foreground group, and the inferences on the group's mass, are completely consistent with this configuration resulting in a high magnification of the $z\sim8$ LBG. In fact, the foreground group would need a peculiarly low mass given its luminosity for there to be low magnification ($\mu\lesssim1.4$) of the LBG.\\

\section{Conclusion}
\label{section:conclusion}
\indent In this paper, we use broad-band HST photometry to identify a candidate galaxy-scale gravitational lensing system composed of a $z\sim8$ LBG source and a small foreground group. To follow-up the system we have obtained very deep VLT/X-Shooter spectroscopy of the LBG and two members of the group. Our new spectroscopic data confirm that the foreground galaxy pair has a redshift of $z=1.327$ through detection of multiple emission lines. A third possible foreground deflector galaxy, not targeted by spectroscopy, has an extended photometric redshift distribution, with a non-negligible likelihood around $z=1.3$. The X-Shooter spectroscopy also allows us to exclude the LBG candidate as a low-redshift interloper with strong emission lines with high confidence ($70\sigma$ in $H_{160}$). Therefore, due to the small angular separation between the LBG and the foreground galaxies, the foreground galaxy pair's spectroscopic redshift and the ruling out of lower redshift solutions for the LBG from the absence of emission lines, the dropout must have been gravitationally magnified to some degree. In this paper, we carried out a detailed modelling with the goal of quantifying the magnification. We argue that the most likely configuration is that the $z\sim 8$ source has been magnified by $\mu=4.3\pm0.2$, producing a secondary demagnified image on the opposite side of the deflector galaxies. We marginally detect this secondary image (with S/N $\sim 2$) in the band with the deepest imaging, $J_{125}$.\\
%
%
\indent The lens model yields a total group mass of $9.62\pm0.31\times10^{11} M_{\odot}$, which is very close to the mass expected from abundance matching. We have obtained a lensing measurement of the stellar mass of a compact galaxy at $z=1.327$, without being affected by the systematic uncertainties that are associated with the inference of stellar mass from spectral energy distribution using population synthesis models. Assuming the standard dark matter fraction found for lower redshift lenses, the brighter central deflector galaxy has a stellar mass-to-light ratio of $M_{\star}/L_{B} = 2.3^{+0.8}_{-0.6} M_{\odot}/L_{\odot}$ within its effective radius. Overall, the properties of the main lensing galaxy are consistent with observations of `red nuggets' at similar redshifts \citep{daddi2005passively, trujillo2006extremely, buitrago2008size, van2008confirmation}. This represents a leap forward compared to other similar measurements from gravitational lensing, which are limited to $z\lesssim1$ (see Fig. \ref{figure:m2L}).\\
\indent Using this interpretation of the data, we find a strongly lensed fraction of galaxies bright than $\approx M_{\star}$ in BoRG to be $10$ per cent. While the uncertainty on this number is large due to the rarity of bright $z\sim8$ objects, this value is quantitatively consistent with the findings of \citet{mason2015correcting}, who found an expected strongly lensed fraction of $3-15$ per cent in the BoRG fields, the recent analysis of the CANDELS fields and the XDF by \citet{barone2015impact} and theoretical modelling by \citet{wyithe2011distortion}. Since \citet{wyithe2011distortion} showed that the effect of magnification bias on the galaxy luminosity function increases with redshift, quantifying and verifying its impact at $z\sim8$ is important for predictions for future surveys targeted at finding $L\gtrsim L_{\star}$ galaxies at yet higher redshift, such as \textit{Euclid} and surveys using \textit{JWST} and \textit{WFIRST}.\\
%
\indent Overall, the strong lensing configuration derived by our work is the most natural interpretation, but a more precise measurement of the magnification of the source and the counter image would allow more robust inferences about the system. Intriguingly, our modelling makes specific predictions testable with deeper HST imaging, including the precise flux ratio between the two images and the elongation of the brighter image. Improving the S/N of the observations by a factor of two would allow both these measurements and hence allow a better determination of the magnification of this $z\sim8$ LBG. Such a goal is within reach of HST as it would require only $\sim10$ orbits of observations.\\

\textbf{Acknowledgments} This work is partially supported by programs number HST/GO 11700, 12587, 12905, and 13767 from NASA through a grant from the Space Telescope Science Institute, which is operated by the Association of Universities for Research in Astronomy, Inc, under NASA contract NAS5-26555.  JSBW acknowledges the support of an Australian Research Council Laureate Fellowship. MT acknowledges the support of an Australia Research Council Future Fellowship. We would like to acknowledge the anonymous referee for their helpful comments, Sharon Rapoport for her initial lens modelling of the system and Stephanie Bernard for her help with the BoRG data.

\bibliographystyle{mn2e}
\bibliography{z8lens}

\begin{thebibliography}{83}
\expandafter\ifx\csname natexlab\endcsname\relax\def\natexlab#1{#1}\fi

\bibitem[{Ade {et~al}\mbox{.}(2013)Ade, Aghanim, Armitage-Caplan, Arnaud,
  Ashdown, Atrio-Barandela, Aumont, Baccigalupi, Banday, Barreiro,
  {et~al.}}]{ade2013planck}
Ade P. {et~al.}, 2013, arXiv preprint arXiv:1303.5062

\bibitem[{Alard(2000)}]{alard2000image}
Alard C., 2000, Astronomy and Astrophysics Supplement Series, 144, 363

\bibitem[{Atek {et~al}\mbox{.}(2011)Atek, Siana, Scarlata, Malkan, McCarthy,
  Teplitz, Henry, Colbert, Bridge, Bunker, {et~al.}}]{atek2011very}
Atek H. {et~al.}, 2011, The Astrophysical Journal, 743, 121

\bibitem[{Auger {et~al}\mbox{.}(2010{\natexlab{a}})Auger, Treu, Bolton,
  Gavazzi, Koopmans, Marshall, Moustakas, \& Burles}]{auger2010sloan}
Auger M., Treu T., Bolton A., Gavazzi R., Koopmans L., Marshall P., Moustakas
  L., Burles S., 2010{\natexlab{a}}, The Astrophysical Journal, 724, 511

\bibitem[{Auger {et~al}\mbox{.}(2010{\natexlab{b}})Auger, Treu, Gavazzi,
  Bolton, Koopmans, \& Marshall}]{auger2010dark}
Auger M., Treu T., Gavazzi R., Bolton A., Koopmans L., Marshall P.,
  2010{\natexlab{b}}, The Astrophysical Journal Letters, 721, L163

\bibitem[{Barkana \& Loeb(2000)}]{barkana2000high}
Barkana R., Loeb A., 2000, The Astrophysical Journal, 531, 613

\bibitem[{Barone-Nugent {et~al}\mbox{.}(2014)Barone-Nugent, Trenti, Wyithe,
  Bouwens, Oesch, Illingworth, Carollo, Su, Stiavelli, Labbe,
  {et~al.}}]{barone2014measurement}
Barone-Nugent R. {et~al.}, 2014, The Astrophysical Journal, 793, 17

\bibitem[{Barone-Nugent {et~al}\mbox{.}(2015)Barone-Nugent, Wyithe, Trenti,
  Treu, Oesch, Bouwens, \& Schmidt}]{barone2015impact}
Barone-Nugent R., Wyithe J., Trenti M., Treu T., Oesch P., Bouwens R., Schmidt
  K.~B., 2015, Monthly Notices of the Royal Astronomical Society

\bibitem[{Benitez(2000)}]{benitez2000bayesian}
Benitez N., 2000, The Astrophysical Journal, 536, 571

\bibitem[{Benitez(2008)}]{benitez2008bayesian}
Benitez N., 2008, The Astrophysical Journal, 536, 571

\bibitem[{Ben{\'i}tez {et~al}\mbox{.}(2004)Ben{\'i}tez, Ford, Bouwens,
  Menanteau, Blakeslee, Gronwall, Illingworth, Meurer, Broadhurst, Clampin,
  {et~al.}}]{benitez2004faint}
Ben{\'i}tez N. {et~al.}, 2004, The Astrophysical Journal Supplement Series,
  150, 1

\bibitem[{Bouwens {et~al}\mbox{.}(2011)Bouwens, Illingworth, Oesch, Labb{\'e},
  Trenti, van Dokkum, Franx, Stiavelli, Carollo, Magee,
  {et~al.}}]{bouwens2011ultraviolet}
Bouwens R. {et~al.}, 2011, The Astrophysical Journal, 737, 90

\bibitem[{Bouwens {et~al}\mbox{.}(2015)Bouwens, Illingworth, Oesch, Trenti,
  Labb{\'e}, Bradley, Carollo, van Dokkum, Gonzalez, Holwerda,
  {et~al.}}]{bouwens2015uv}
Bouwens R. {et~al.}, 2015, The Astrophysical Journal, 803, 34

\bibitem[{Bouwens {et~al}\mbox{.}(2008)Bouwens, Illingworth, Franx, \&
  Ford}]{bouwens2008z}
Bouwens R.~J., Illingworth G.~D., Franx M., Ford H., 2008, The Astrophysical
  Journal, 686, 230

\bibitem[{Bradley {et~al}\mbox{.}(2012)Bradley, Trenti, Oesch, Stiavelli, Treu,
  Bouwens, Shull, Holwerda, \& Pirzkal}]{bradley2012brightest}
Bradley L. {et~al.}, 2012, The Astrophysical Journal, 760, 108

\bibitem[{Buitrago {et~al}\mbox{.}(2008)Buitrago, Trujillo, Conselice, Bouwens,
  Dickinson, \& Yan}]{buitrago2008size}
Buitrago F., Trujillo I., Conselice C.~J., Bouwens R.~J., Dickinson M., Yan H.,
  2008, The Astrophysical Journal Letters, 687, L61

\bibitem[{{Bundy} {et~al}\mbox{.}(2007){Bundy}, {Treu}, \&
  {Ellis}}]{bundy2007mass}
{Bundy} K., {Treu} T., {Ellis} R.~S., 2007, The Astrophysical Journal Letters,
  665, L5

\bibitem[{{Cappellari} {et~al}\mbox{.}(2009){Cappellari}, {di Serego
  Alighieri}, {Cimatti}, {Daddi}, {Renzini}, {Kurk}, {Cassata}, {Dickinson},
  {Franceschini}, {Mignoli}, {Pozzetti}, {Rodighiero}, {Rosati}, \&
  {Zamorani}}]{cappellari2009dynamical}
{Cappellari} M. {et~al.}, 2009, The Astrophysical Journal Letters, 704, L34

\bibitem[{Castellano {et~al}\mbox{.}(2010)Castellano, Fontana, Paris, Grazian,
  Pentericci, Boutsia, Santini, Testa, Dickinson, Giavalisco,
  {et~al.}}]{castellano2010bright}
Castellano M. {et~al.}, 2010, arXiv preprint arXiv:1007.5396

\bibitem[{Coe {et~al}\mbox{.}(2007)Coe, Ben{\'i}tez, S{\'a}nchez, Jee, Bouwens,
  \& Ford}]{coe2007galaxies}
Coe D., Ben{\'i}tez N., S{\'a}nchez S., Jee M., Bouwens R., Ford H., 2007, The
  Astronomical Journal, 132, 926

\bibitem[{Comerford {et~al}\mbox{.}(2002)Comerford, Haiman, \&
  Schaye}]{comerford2002constraining}
Comerford J.~M., Haiman Z., Schaye J., 2002, The Astrophysical Journal, 580, 63

\bibitem[{Curtis-Lake {et~al}\mbox{.}(2014)Curtis-Lake, McLure, Dunlop, Rogers,
  Targett, Dekel, Ellis, Faber, Ferguson, Grogin, {et~al.}}]{curtis2014no}
Curtis-Lake E. {et~al.}, 2014, arXiv preprint arXiv:1409.1832

\bibitem[{Daddi {et~al}\mbox{.}(2005)Daddi, Renzini, Pirzkal, Cimatti,
  Malhotra, Stiavelli, Xu, Pasquali, Rhoads, Brusa,
  {et~al.}}]{daddi2005passively}
Daddi E. {et~al.}, 2005, The Astrophysical Journal, 626, 680

\bibitem[{Damjanov {et~al}\mbox{.}(2009)Damjanov, McCarthy, Abraham,
  Glazebrook, Yan, Mentuch, {Le Borgne}, Savaglio, Crampton, Murowinski,
  {et~al.}}]{damjanov2009red}
Damjanov I. {et~al.}, 2009, The Astrophysical Journal, 695, 101

\bibitem[{{Dutton} {et~al}\mbox{.}(2010){Dutton}, {Conroy}, {van den Bosch},
  {Prada}, \& {More}}]{dutton2010kinematic}
{Dutton} A.~A., {Conroy} C., {van den Bosch} F.~C., {Prada} F., {More} S.,
  2010, Monthly Notices of the Royal Astronomical Society, 407, 2

\bibitem[{Faber \& Jackson(1976)}]{faber1976velocity}
Faber S., Jackson R.~E., 1976, The Astrophysical Journal, 204, 668

\bibitem[{Fialkov \& Loeb(2015)}]{fialkov2015distortion}
Fialkov A., Loeb A., 2015, arXiv preprint arXiv:1502.03141

\bibitem[{Finkelstein {et~al}\mbox{.}(2013)Finkelstein, Papovich, Dickinson,
  Song, Tilvi, Koekemoer, Finkelstein, Mobasher, Ferguson, Giavalisco,
  {et~al.}}]{finkelstein2013galaxy}
Finkelstein S. {et~al.}, 2013, Nature, 502, 524

\bibitem[{Finkelstein {et~al}\mbox{.}(2012)Finkelstein, Papovich, Ryan, Pawlik,
  Dickinson, Ferguson, Finlator, Koekemoer, Giavalisco, Cooray,
  {et~al.}}]{finkelstein2012candels}
Finkelstein S.~L. {et~al.}, 2012, The Astrophysical Journal, 758, 93

\bibitem[{Finkelstein {et~al}\mbox{.}(2014)Finkelstein, {Ryan Jr}, Papovich,
  Dickinson, Song, Somerville, Ferguson, Salmon, Giavalisco, Koekemoer,
  {et~al.}}]{finkelstein2014evolution}
Finkelstein S.~L. {et~al.}, 2014, arXiv preprint arXiv:1410.5439

\bibitem[{Giavalisco(2002)}]{giavalisco2002lyman}
Giavalisco M., 2002, Annual Review of Astronomy and Astrophysics, 40, 579

\bibitem[{Grazian {et~al}\mbox{.}(2012)Grazian, Castellano, Fontana,
  Pentericci, Dunlop, McLure, Koekemoer, Dickinson, Faber, Ferguson,
  {et~al.}}]{grazian2012size}
Grazian A. {et~al.}, 2012, Astronomy \& Astrophysics

\bibitem[{Grogin {et~al}\mbox{.}(2011)Grogin, Kocevski, Faber, Ferguson,
  Koekemoer, Riess, Acquaviva, Alexander, Almaini, Ashby,
  {et~al.}}]{grogin2011candels}
Grogin N.~A. {et~al.}, 2011, The Astrophysical Journal Supplement Series, 197,
  35

\bibitem[{Illingworth {et~al}\mbox{.}(2013)Illingworth, Magee, Oesch, Bouwens,
  Labb{\'e}, Stiavelli, van Dokkum, Franx, Trenti, Carollo,
  {et~al.}}]{illingworth2013hst}
Illingworth G. {et~al.}, 2013, The Astrophysical Journal Supplement Series,
  209, 6

\bibitem[{Keeton(2001{\natexlab{a}})}]{keeton2001catalog}
Keeton C., 2001{\natexlab{a}}, arXiv preprint astro-ph/0102341

\bibitem[{Keeton(2001{\natexlab{b}})}]{keeton2001computational}
Keeton C., 2001{\natexlab{b}}, Arxiv preprint astro-ph/0102340

\bibitem[{Khochfar {et~al}\mbox{.}(2007)Khochfar, Silk, Windhorst, \& {Ryan
  Jr}}]{khochfar2007evolving}
Khochfar S., Silk J., Windhorst R., {Ryan Jr} R., 2007, The Astrophysical
  Journal Letters, 668, L115

\bibitem[{{Kochanek}(1995)}]{kochanek1995evidence}
{Kochanek} C.~S., 1995, The Astrophysical Journal, 445, 559

\bibitem[{Koekemoer {et~al}\mbox{.}(2011)Koekemoer, Faber, Ferguson, Grogin,
  Kocevski, Koo, Lai, Lotz, Lucas, McGrath, {et~al.}}]{koekemoer2011candels}
Koekemoer A.~M. {et~al.}, 2011, The Astrophysical Journal Supplement Series,
  197, 36

\bibitem[{Komatsu {et~al}\mbox{.}(2010)Komatsu, Smith, Dunkley, Bennett, Gold,
  Hinshaw, Jarosik, Larson, Nolta, Page, {et~al.}}]{komatsu2010seven}
Komatsu E. {et~al.}, 2010, arXiv preprint arXiv:1001.4538

\bibitem[{Koopmans {et~al}\mbox{.}(2009)Koopmans, Bolton, Treu, Czoske, Auger,
  Barnab{\`e}, Vegetti, Gavazzi, Moustakas, \& Burles}]{koopmans2009structure}
Koopmans L. {et~al.}, 2009, The Astrophysical Journal Letters, 703, L51

\bibitem[{Mason {et~al}\mbox{.}(2015)Mason, Treu, Schmidt, Collett, Trenti,
  Marshall, Barone-Nugent, Bradley, Stiavelli, \& Wyithe}]{mason2015correcting}
Mason C. {et~al.}, 2015

\bibitem[{McLure {et~al}\mbox{.}(2013)McLure, Dunlop, Bowler, Curtis-Lake,
  Schenker, Ellis, Robertson, Koekemoer, Rogers, Ono, {et~al.}}]{mclure2013new}
McLure R. {et~al.}, 2013, Monthly Notices of the Royal Astronomical Society,
  stt627

\bibitem[{Modigliani {et~al}\mbox{.}(2010)Modigliani, Goldoni, Royer, Haigron,
  Guglielmi, Fran\c{c}ois, Horrobin, Bristow, Vernet, Moehler,
  {et~al.}}]{modigliani2010x}
Modigliani A. {et~al.}, 2010, in {SPIE Astronomical Telescopes+
  Instrumentation}, International Society for Optics and Photonics, pp.
  773728--773728

\bibitem[{Mu{\~n}oz \& Loeb(2008)}]{munoz2008verifying}
Mu{\~n}oz J.~A., Loeb A., 2008, Monthly Notices of the Royal Astronomical
  Society, 385, 2175

\bibitem[{Navarro {et~al}\mbox{.}(1997)Navarro, Frenk, \&
  White}]{navarro1997universal}
Navarro J.~F., Frenk C.~S., White S.~D., 1997, The Astrophysical Journal, 490,
  493

\bibitem[{Newman {et~al}\mbox{.}(2012)Newman, Ellis, Bundy, \&
  Treu}]{newman2012can}
Newman A.~B., Ellis R.~S., Bundy K., Treu T., 2012, The Astrophysical Journal,
  746, 162

\bibitem[{Newman {et~al}\mbox{.}(2015)Newman, Ellis, \&
  Treu}]{newman2015luminous}
Newman A.~B., Ellis R.~S., Treu T., 2015, arXiv preprint arXiv:1503.05282

\bibitem[{Oesch {et~al}\mbox{.}(2009)Oesch, Bouwens, Carollo, Illingworth,
  Trenti, Stiavelli, Magee, Labb{\'e}, \& Franx}]{oesch2009structure}
Oesch P. {et~al.}, 2009, The Astrophysical Journal Letters, 709, L21

\bibitem[{Oesch {et~al}\mbox{.}(2012)Oesch, Bouwens, Illingworth, Gonzalez,
  Trenti, van Dokkum, Franx, Labb{\'e}, Carollo, \& Magee}]{oesch2012bright}
Oesch P. {et~al.}, 2012, The Astrophysical Journal, 759, 135

\bibitem[{Oesch {et~al}\mbox{.}(2015)Oesch, van Dokkum, Illingworth, Bouwens,
  Momcheva, Holden, Roberts-Borsani, Smit, Franx, Labbe,
  {et~al.}}]{oesch2015spectroscopic}
Oesch P. {et~al.}, 2015, arXiv preprint arXiv:1502.05399

\bibitem[{Oke \& Gunn(1983)}]{oke1983secondary}
Oke J., Gunn J., 1983, The Astrophysical Journal, 266, 713

\bibitem[{Ono {et~al}\mbox{.}(2011)Ono, Ouchi, Mobasher, Dickinson, Penner,
  Shimasaku, Weiner, Kartaltepe, Nakajima, Nayyeri,
  {et~al.}}]{ono2011spectroscopic}
Ono Y. {et~al.}, 2011, The Astrophysical Journal, 744, 83

\bibitem[{Robertson {et~al}\mbox{.}(2015)Robertson, Ellis, Furlanetto, \&
  Dunlop}]{robertson2015cosmic}
Robertson B.~E., Ellis R.~S., Furlanetto S.~R., Dunlop J.~S., 2015, The
  Astrophysical Journal Letters, 802, L19

\bibitem[{Robertson {et~al}\mbox{.}(2013)Robertson, Furlanetto, Schneider,
  Charlot, Ellis, Stark, McLure, Dunlop, Koekemoer, Schenker,
  {et~al.}}]{robertson2013new}
Robertson B.~E. {et~al.}, 2013, The Astrophysical Journal, 768, 71

\bibitem[{Ruff {et~al}\mbox{.}(2011)Ruff, Gavazzi, Marshall, Treu, Auger, \&
  Brault}]{ruff2011sl2s}
Ruff A., Gavazzi R., Marshall P., Treu T., Auger M., Brault F., 2011, The
  Astrophysical Journal, 727, 96

\bibitem[{Schenker {et~al}\mbox{.}(2013)Schenker, Robertson, Ellis, Ono,
  McLure, Dunlop, Koekemoer, Bowler, Ouchi, Curtis-Lake,
  {et~al.}}]{schenker2013uv}
Schenker M.~A. {et~al.}, 2013, The Astrophysical Journal, 768, 196

\bibitem[{Schmidt {et~al}\mbox{.}(2014)Schmidt, Treu, Trenti, Bradley, Kelly,
  Oesch, Holwerda, Shull, \& Stiavelli}]{schmidt2014luminosity}
Schmidt K.~B. {et~al.}, 2014, The Astrophysical Journal, 786, 196

\bibitem[{S{\'e}rsic(1963)}]{sersic1963influence}
S{\'e}rsic J., 1963, Boletin de la Asociacion Argentina de Astronomia La Plata
  Argentina, 6, 41

\bibitem[{Shibuya {et~al}\mbox{.}(2015)Shibuya, Ouchi, \&
  Harikane}]{shibuya2015morphologies}
Shibuya T., Ouchi M., Harikane Y., 2015, arXiv preprint arXiv:1503.07481

\bibitem[{Shull {et~al}\mbox{.}(2012)Shull, Harness, Trenti, \&
  Smith}]{shull2012critical}
Shull J., Harness A., Trenti M., Smith B., 2012, The Astrophysical Journal,
  747, 100

\bibitem[{Stark {et~al}\mbox{.}(2015{\natexlab{a}})Stark, Richard, Charlot,
  Cl{\'e}ment, Ellis, Siana, Robertson, Schenker, Gutkin, \&
  Wofford}]{stark2015spectroscopic2}
Stark D.~P. {et~al.}, 2015{\natexlab{a}}, Monthly Notices of the Royal
  Astronomical Society, 450, 1846

\bibitem[{Stark {et~al}\mbox{.}(2015{\natexlab{b}})Stark, Walth, Charlot,
  Clement, Feltre, Gutkin, Richard, Mainali, Robertson, Siana,
  {et~al.}}]{stark2015spectroscopic}
Stark D.~P. {et~al.}, 2015{\natexlab{b}}, arXiv preprint arXiv:1504.06881

\bibitem[{Steidel {et~al}\mbox{.}(1996)Steidel, Giavalisco, Dickinson, \&
  Adelberger}]{steidel1996spectroscopy}
Steidel C., Giavalisco M., Dickinson M., Adelberger K., 1996, Arxiv preprint
  astro-ph/9604140

\bibitem[{Trenti {et~al}\mbox{.}(2011)Trenti, Bradley, Stiavelli, Oesch, Treu,
  Bouwens, Shull, MacKenty, Carollo, \& Illingworth}]{trenti2011brightest}
Trenti M. {et~al.}, 2011, The Astrophysical Journal Letters, 727, L39

\bibitem[{Trenti {et~al}\mbox{.}(2012)Trenti, Bradley, Stiavelli, Shull, Oesch,
  Bouwens, Mu{\~n}oz, Romano-Diaz, Treu, Shlosman,
  {et~al.}}]{trenti2012overdensities}
Trenti M. {et~al.}, 2012, The Astrophysical Journal, 746, 55

\bibitem[{Trenti {et~al}\mbox{.}(2014)Trenti, Perna, \&
  Jimenez}]{trenti2014luminosity}
Trenti M., Perna R., Jimenez R., 2014, arXiv preprint arXiv:1406.1503

\bibitem[{Trenti {et~al}\mbox{.}(2010)Trenti, Smith, Hallman, Skillman, \&
  Shull}]{trenti2010well}
Trenti M., Smith B.~D., Hallman E.~J., Skillman S.~W., Shull J.~M., 2010, The
  Astrophysical Journal, 711, 1198

\bibitem[{Trenti \& Stiavelli(2008)}]{trenti2008cosmic}
Trenti M., Stiavelli M., 2008, The Astrophysical Journal, 676, 767

\bibitem[{{Treu}(2010)}]{treu2010strong}
{Treu} T., 2010, Annual Review of Astronomy and Astrophysics, 48, 87

\bibitem[{Treu \& Koopmans(2002)}]{treu2002internal}
Treu T., Koopmans L., 2002, The Astrophysical Journal, 575, 87

\bibitem[{{Treu} \& {Koopmans}(2004)}]{treu2004massive}
{Treu} T., {Koopmans} L.~V.~E., 2004, The Astrophysical Journal, 611, 739

\bibitem[{Treu {et~al}\mbox{.}(2013)Treu, Schmidt, Trenti, Bradley, \&
  Stiavelli}]{treu2013changing}
Treu T., Schmidt K.~B., Trenti M., Bradley L.~D., Stiavelli M., 2013, arXiv
  preprint arXiv:1308.5985

\bibitem[{Treu {et~al}\mbox{.}(2002)Treu, Stiavelli, Casertano, M{\o}ller, \&
  Bertin}]{treu2002evolution}
Treu T., Stiavelli M., Casertano S., M{\o}ller P., Bertin G., 2002, The
  Astrophysical Journal Letters, 564, L13

\bibitem[{Trujillo {et~al}\mbox{.}(2006)Trujillo, Feulner, Goranova, Hopp,
  Longhetti, Saracco, Bender, Braito, {Della Ceca}, Drory,
  {et~al.}}]{trujillo2006extremely}
Trujillo I. {et~al.}, 2006, Monthly Notices of the Royal Astronomical Society:
  Letters, 373, L36

\bibitem[{{Van Dokkum} {et~al}\mbox{.}(2008){Van Dokkum}, Franx, Kriek, Holden,
  Illingworth, Magee, Bouwens, Marchesini, Quadri, Rudnick,
  {et~al.}}]{van2008confirmation}
{Van Dokkum} P.~G. {et~al.}, 2008, The Astrophysical Journal Letters, 677, L5

\bibitem[{{Wallington} \& {Narayan}(1993)}]{wallington1993influence}
{Wallington} S., {Narayan} R., 1993, The Astrophysical Journal, 403, 517

\bibitem[{Wayth \& Webster(2006)}]{wayth2006lensview}
Wayth R., Webster R., 2006, Monthly Notices of the Royal Astronomical Society,
  372, 1187

\bibitem[{Wong {et~al}\mbox{.}(2012)Wong, Ammons, Keeton, \&
  Zabludoff}]{wong2012optimal}
Wong K.~C., Ammons S.~M., Keeton C.~R., Zabludoff A.~I., 2012, The
  Astrophysical Journal, 752, 104

\bibitem[{Wong {et~al}\mbox{.}(2014)Wong, Tran, Suyu, Momcheva, Brammer,
  Brodwin, Gonzalez, Halkola, Kacprzak, Koekemoer,
  {et~al.}}]{wong2014discovery}
Wong K.~C. {et~al.}, 2014, The Astrophysical Journal Letters, 789, L31

\bibitem[{Wyithe {et~al}\mbox{.}(2011)Wyithe, Yan, Windhorst, \&
  Mao}]{wyithe2011distortion}
Wyithe J., Yan H., Windhorst R., Mao S., 2011, Nature, 469, 181

\bibitem[{Wyithe \& Loeb(2011)}]{wyithe2011extrapolating}
Wyithe J. S.~B., Loeb A., 2011, Monthly Notices of the Royal Astronomical
  Society: Letters, 413, L38

\bibitem[{Yan {et~al}\mbox{.}(2011)Yan, Yan, Zamojski, Windhorst, McCarthy,
  Fan, R{\"o}ttgering, Koekemoer, Robertson, Dav{\'e},
  {et~al.}}]{yan2011probing}
Yan H. {et~al.}, 2011, The Astrophysical Journal Letters, 728, L22

\end{thebibliography}

\label{lastpage}
\end{document}